\documentclass{webofc}
\usepackage[varg]{txfonts}   
\begin{document}
\title{Thermal Convection in Stars and in Their Atmosphere}
%
%

\author{\firstname{Friedrich} \lastname{Kupka}\inst{1,2}\fnsep\thanks{\email{<friedrich.kupka@uni-goettingen.de>,
             <kupka@mps.mpg.de>}} }

\institute{formerly at: Institute for Astrophysics, Georg-August-University G\"ottingen, \\ Friedrich-Hund-Platz 1,
             D-37077 G\"ottingen, Germany
             \and
             MPI for Solar System Research, Justus-von-Liebig-Weg 3, D-37077 G\"ottingen, Germany
             }

\abstract{Thermal convection is one of the main mechanisms of heat transport and mixing
              in stars in general and also in the photospheric layers which emit the radiation that we observe 
              with astronomical instruments. The present lecture notes first introduce the role of convection in astrophysics
              and explain the basic physics of convection. This is followed by an overview on the modelling of convection.
              Challenges and pitfalls in numerical simulation based modelling are discussed subsequently. Finally, 
              a particular application for the previously introduced concepts is described in more detail: 
              the study of convective overshooting into stably stratified layers around convection zones in stars.
} 
\maketitle

\section{Introduction}  \label{intro}

Convection is an important mechanism for energy transport and mixing in stars.
Both its analytical and its numerical modelling are particularly challenging and 
have remained key topics in stellar physics ever since it had been realized that stars 
can have convectively unstable regions at their surface and in their interior 
(\cite{schwarzschild06r,unsoeld30r,biermann32r,siedentopf33r,siedentopf35r}).

This review article provides an introduction into the subject, but 
has to neglect some important varieties of convection: magneto-convection,
double-diffusive convection in multicomponent fluids, and, mostly,
also convection in (rapidly) rotating objects. Those subjects are covered by 
further articles in this book. Including them would have been prohibitive for 
keeping this exposition within reasonable limits. For completeness though
some references on these subjects are also provided just below.

Concepts from fluid mechanics are essential in the study of convection. 
Readers interested in an introduction to fluid dynamics as a supplement to this
review are referred to \cite{batchelor00b,landau63b}. Those introductions omit 
the subject of magnetohydrodynamics (MHD) which is covered, for instance, in 
\cite{landau84b} or also in \cite{galtier16b}, who provides a modern exposition of 
MHD. All these books are useful for finding further introductory texts
on their main topics, too. Concerning the physics of turbulent flows we refer
in particular to \cite{pope00b} for an introduction to statistical concepts,
one-point closures, and large eddy simulations, as well as to 
\cite{lesieur08b} for an introduction to two-point closures and other
techniques not covered by the previous reference, such as 2D turbulence
and turbulence in geophysics. These topics are also of interest to readers
specializing in astrophysics. A critical review of the different concepts used 
in studying the physics of turbulence can be found in \cite{tsinober09b}. 

There is a vast mathematical literature about the foundations of numerical 
methods for solving the partial differential equations of fluid dynamics which 
are at the core of theoretical studies of convection as well as on 
strategies concerning their implementation in a numerical simulation code. For 
beginners the introduction by \cite{ferziger13b} is particularly useful. It focusses 
on finite volume methods and on incompressible flows but it also provides 
a discussion of more general methods and the case of compressible flows. 
The reader interested in the basics and implementations of finite difference methods 
may find both modern introductions (\cite{strikwerda89b} and later re-editions)
as well as classical books such as \cite{richtmyer67b} useful. As a standard 
reference for spectral methods, popular in numerical models of stellar convection, 
the books of \cite{ccanuto10b,ccanuto14b} may be consulted.
Finally, a more mathematical account that includes the case of finite
element methods but also the basics of modern shock resolving methods
is given in \cite{quarteroni94b}. These books should help
in finding further literature on this still rapidly evolving subject.

As is discussed below 2D and 3D numerical approximations to
the basic equations of fluid dynamics are too expensive for
a direct application to stellar evolution calculations. Classical stellar 
structure and evolution models are thus still rooted in computationally
much less demanding, semi-analytical models. In astrophysics the most 
frequently used  among them is mixing length theory (\cite{biermann32r,bv58b}).
A detailed account of this approach is given in classical texts
on the theory of stellar structure and evolution such as \cite{cox04b}.

For specialized subtopics concerning convection in stars the comprehensive and 
freely accessible reviews of the Living Reviews series on Solar Physics
and on Computational Astrophysics are a good starting point to find further information. 
Subjects dealt with by these series include the large scale dynamics of the solar 
convection zone and tachocline \cite{miesch05b}, an account of solar surface convection 
for the case where magnetic fields are neglected \cite{nordlund09r}, the problem of
interaction between convection and pulsation \cite{houdek15b}, and the modelling
of stellar convection with both semi-analytical and numerical methods \cite{kupka17b}.
For advanced subjects such as supergranulation \cite{rincon18b} specialized reviews
are available as well. This also holds for extensions concerning magnetic fields and
the solar cycle, for example, dynamo models of the solar cycle \cite{charbonneau10b},
solar surface magneto-convection \cite{stein12b}, or modelling of magnetism and 
dynamo action for both the solar and the stellar case (\cite{brun17b}).

This introduction hence limits itself to the following topics: Sect.~\ref{sec-2} gives an 
overview on the role of convection in astrophysics and on the basic physics of convection. 
Sect.~\ref{sec-3} deals with the modelling of convection, with a focus on semi-analytical 
models. Sect.~\ref{sec-4} provides a discussion of some challenges and pitfalls in 
numerical simulations of convection. Sect.~\ref{sec-5} discusses the problem
of calculating the extent of overshooting of convection into 
neighbouring, locally stably stratified regions, as a challenging example
for the different modelling techniques introduced in the previous sections.
Sect.~\ref{sec-6} provides a summary of this material.

\section{Convection in astrophysics and the basic physics of convection}  \label{sec-2}
\subsection{The physics of convection} \label{sec-2.1}

Convection is caused by a hydrodynamical instability which can occur in a fluid layer 
stratified due to gravitational force and subject to a temperature difference between 
the ``top'' and the ``bottom'' of that layer. Gravity specifies this distinct ``vertical'' 
direction as the one aligned with its action. In a sufficiently slowly rotating and hence 
spherically symmetric star this direction of course coincides with the radial one. Thus, 
the top of such a layer is located towards the surface of a star, its bottom towards the 
centre. These definitions are also evident for laboratory models such as convection 
occurring between two horizontally mounted 
plates.\footnote{This setup can be generalized by including other forces
                        or by letting an electrical force take over the role of gravitation,
                        but we focus on the standard case here and in the following.}
Consider a fluid stratified such that it is initially in hydrostatic equilibrium
and $\rho_{\rm top} < \rho_{\rm bottom}$. Linear stability analysis demonstrates that
this configuration can become unstable depending on the distribution of temperature
along the vertical direction. As shown in Sect.~\ref{sec-2.4}, if we consider adiabatic 
expansion (without viscous friction) of a fluid volume that is perturbed (considered 
moved away) vertically from its initial position such that through this expansion it 
attains the mean pressure of its new environment again, it is then crucial whether this 
displaced fluid has a lower density than its new environment. If that is the case 
a net buoyancy force prevails and the fluid is unstable to convection.

In practice, at least in a stellar context small perturbations are always present 
and they will initiate such a convective instability. This leads to velocity fields 
building up with time. The evolution of this process can be very accurately predicted 
by the conservation laws of hydrodynamics: the Navier-Stokes equation (NSE) which
ensures conservation of momentum, and its associated conservation laws
for mass (continuity equation), and energy. Eventually, this causes heat transport
in the fluid and mixing of the fluid. The convectively driven velocity field also
couples to pulsation and shear flows in stars induced by pulsational instabilities
and rotation. Since the fluid in stars is actually a plasma, it is generally
magnetic and its velocity field can hence either cause or react to magnetic
fields as well.

The key criterion for convective instability in stars was originally suggested by 
\cite{schwarzschild06r} and is also derived in Sect.~\ref{sec-2.4}. The Schwarzschild
criterion assures that a given stratification is unstable to convection, if the local 
(horizontally averaged) temperature gradient is steeper than the adiabatic one.
With the definition of the dimensionless temperature gradient $\nabla := (d\ln T/d\ln P)$
this can be written as $\nabla > \nabla_{\rm ad}$. In turn, the gradient of radiative diffusion
for a spherically symmetric star in hydrostatic equilibrium (\cite{cox04b}) can be written
as $\nabla_{\rm rad} = (3 \kappa_{\rm ross}\,P\,L_r) / (16\pi a c G T^4 M_r)$. Here,
$P$ is the pressure, $L_r$ the luminosity at radial coordinate $r$, $M_r$ is the mass inside 
a radius of $r$, $T$ is the temperature, and $\kappa_{\rm ross}$ is the Rosseland opacity.
For a first analysis it suffices to compare $\nabla_{\rm rad}$ and $\nabla_{\rm ad}$ with each
other to find the main sources of convective instability. The first one was identified to be partial 
ionization (by \cite{unsoeld30r} for the zone of partial ionization of hydrogen in the Sun). Basically, 
in cool stars of spectral type A to M the quantity $\nabla_{\rm ad}$ drops from $\sim 0.4$ down to 
$0.05 \dots 0.1$ in this zone whence $\nabla_{\rm ad} < \nabla_{\rm rad}$ (see also \cite{cox04b}). An even 
more important source is triggered simultaneously in this zone: high opacity due to partial ionization of hydrogen, 
but in general also of helium or ``iron peak'' elements. The former ones are important for A to M type stars. 
Iron peak opacity is important for hot stars (O and B type, but also A type stars when radiative diffusion is
accounted for, see \cite{stothers00b,richer00b}). Again, $\nabla_{\rm ad} < \nabla_{\rm rad}$ in this case, 
but this time primarily due to a large $\nabla_{\rm rad}$ instead of a small $\nabla_{\rm ad}$. The third main 
reason for convection in stars is high luminosity by efficient energy generation from nuclear fusion. 
There, $\varepsilon_c = d L_r / d M_r \sim L_r / M_r$ for small $M_r$ which is large for energy production 
dominated by the CNO cycle and all later burning stages. Thus, along the main sequence convective cores 
occur from O down to F type stars. There, $\nabla_{\rm ad} < \nabla_{\rm rad}$ because of a large 
$\nabla_{\rm rad}$. A steep temperature gradient hence triggers the convective instability and the first model 
for this process in astrophysics was proposed by \cite{biermann32r}.

\subsection{Examples from astrophysics and geophysics} \label{sec-2.2} 

On Earth convection can occur both in the ocean (or lakes, etc.) and in the 
atmosphere and in a different physical parameter range also in its interior.
In the atmosphere of the Earth convection may be caused by the specific
moisture content of air, by the Sun heating the surface during the day,
or by wind moving cold air, for instance from above the arctic sea ice,
over warmer, open ocean water (see \cite{hartmann97b} for images
of this scenario). The latter two scenarios are known under the name of
the convective planetary boundary layer (CBL/PBL). This is a local phenomenon:
the lowermost part of the atmosphere of the Earth can be convectively
unstable in some area, but stable elsewhere and this even changes as
function of time for each area. That is different from the solar or stellar 
case, where the convective instability usually occurs at all locations 
at a given radius (apart from special local conditions caused by magnetic 
features such as sun spots and the like). 

In astrophysics solar granulation is the most well-known observational
feature caused by convection (see \cite{schwarzschild59b} and \cite{leighton63b}
for some early photographic images, more recent images can be found in the 
reviews mentioned in the introduction). Its characteristic structure is
caused by hot upflows which are separated from each other by regions of cold
downflows. This is quite different from the hot and narrow upwards flowing
plumes observed in the convective planetary boundary layer of the atmosphere
of the Earth. Clearly, the boundary conditions of each system are important,
i.e., whether the system becomes convective due to a heating source
at the bottom (PBL), a cooling layer on top (Sun, convection in the ocean),
or a mixture of both (other PBL scenarios), determines the flow topology.
Extensive studies on this subject have been performed in the atmospheric 
sciences (\cite{wyngaard87b,moeng89b,moeng90b,wyngaard91b,piper95b}), but 
see also \cite{stein98b} for the astrophysical case. The astrophysical and geophysical 
scenarios for which convection occurs are nevertheless sufficiently closely related 
to each other such that studies of results on this subject in the neighbouring field 
are highly worthwhile for either side.

\subsection{Astrophysical implications} \label{sec-2.3}

On a phenomenological level one can characterize the effects of convection
on the physical properties of stars as follows. Through its action on temperature 
gradients and through the inhomogeneity of physical quantities such as temperature 
it causes (for instance, at stellar surfaces through granulation), convective heat transfer 
modifies the emitted radiation of stellar atmospheres and thus it changes the
photometric colours of stars (see also \cite{smalley97b}), the profiles of absorption lines generated 
in their photosphere (cf.\ \cite{heiter02b}, or \cite{nordlund09r} for a summary), and the 
chromospheric activity of stars. From a more general perspective this leads to uncertainties 
in secondary distance indicators when deriving stellar parameters from 
analyzing the atmosphere of a star to determine its absolute magnitude and hence its distance.

Stellar structure and stellar evolution are influenced by convection especially
along the pre-main sequence and during post-main sequence evolution.
In typical calibrations convection contributes to the uncertainties in determining
the effective temperature for pre-main sequence models up to $T_{\rm eff} \pm 175\,\rm K$
(e.g.\ \cite{montalban04b}). Along the main sequence it influences stellar radii through the 
size of the convective envelope of F to M type stars and particularly the stellar luminosity 
at the end of the main sequence. For the lower red giant branch (RGB) for stars with $1 M_{\odot}$ 
evolutionary tracks are subject to an additional uncertainty of $T_{\rm eff} \pm 100\,\rm K$
(e.g.\ \cite{montalban04b}). Different convection models and calibrations based on 3D hydrodynamical 
simulations have been proposed to solve this problem (e.g.\ \cite{trampedach99b,trampedach14c}). 
In the end convection influences the mass determination of stars and the interpretation of 
Hertzsprung-Russell diagram (HRD).

Through its capability of overshooting into stable layers convective mixing 
modifies concentration gradients and thus eventually the evolution of convective cores
and hence stellar lifetimes, stellar chemical composition, and stellar (element) yield rates
at the end of stellar evolution (cf.\ the discussion in the introduction of \cite{canuto99b}). 
The effective depth of a convection zone caused by mixing can determine the
destruction of trace elements such as $\rm {}^7 Li$ ($T_{\rm b} \sim 2.5 \cdot 10^6\,\rm K$,
see also \cite{dantona03b}) and likewise also the structure and composition of progenitors 
of core collapse supernovae (cf.\ \cite{canuto99b}), and the production of Li and other 
elements in late stages of stellar evolution.

Convection also couples to mean velocity fields and the magnetic fields of 
stars. It can hence also excite and damp pulsation. Examples include
the computation of frequencies of pressure modes in the Sun and solar-like stars
(\cite{baturin95b,rosenthal99b}) and the excitation and driving of pulsation 
(\cite{stein2001b,samadi06b,belkacem06c})). Convection is thus also investigated 
with the tools provided by helio- and asteroseismology. Its role in angular momentum
transport, the generation of magnetic dynamos, and the origin of long-term cycles
of solar and stellar variability are now increasingly studied not only by observational
means, but also by global numerical simulations of convection, which resolve
the largest scales of the convective zones (e.g., \cite{miesch05b,nelson14b,brun17b}).

\subsection{Local stability analysis for the case of convection} \label{sec-2.4}

The local stability analysis for convection is usually based on the following assumptions:
\begin{enumerate}
\item Consider a fluid element of size $\ell$, 
\item with $\ell$ less than the lengths along which the stellar structure substantially changes:
         $\ell < H_p = -P/(dP/dr)$ and $\ell < R_{*}$, etc., 
\item and assume instantaneous pressure equilibrium of fluid elements with their environment (thus: subsonic flow),
\item furthermore that there is no occurrence of ``acoustic phenomena'' such as stellar pulsation due to sound waves 
        (pressure modes), shock fronts, etc.,
\item with temperature and density fluctuations small against the mean temperature and density at a given height. 
        Assumptions 1 to 5 are essentially the Boussinesq approximation.
\item For stellar applications also assume the viscosity to be very small
\item and consider that we can neglect non-local effects, shear flow and rotation, concentration gradients and magnetic fields. Apart 
        from neglecting non-local effects, assumptions 6 to 7 can also be relaxed as part of more general linear stability analyses.
\end{enumerate}

We now discuss a simple variant of this kind of analysis. Consider a fluid without compositional 
gradient in an external, gravitational field which points towards the negative $z$ direction. 
A fluid element of specific volume $V(P,S)$ were to move adiabatically, i.e., without 
heat exchange with its environment, against the direction of gravity from a height $z$ 
to a height $z+\xi$. There, a pressure $P^{\prime}$ holds and thus after the displacement the specific 
volume of the mass element has changed to $V(P^{\prime}, S)$ while keeping its entropy $S$.

The stratification is stable, i.e., the fluid element is moved back to its
original position $z$, if the fluid element is heavier than the fluid otherwise located 
at $z+\xi$, which has a specific volume $V(P^{\prime}, S^{\prime})$.

The pressure is $P$ at $z$ and $P^{\prime}$ at $z+\xi$, both inside and outside the fluid element. Thus, pressure
equilibrium follows for a low Mach number flow with no acoustic phenomena (assumptions 3+4). The assumption 
of adiabaticity requires small temperature and density fluctuations to hold (that is assumption 5). Generalizations
beyond assumptions 6+7 would have to be introduced at this point, too.

We consider them to hold and hence stability requires $V(P^{\prime}, S^{\prime}) - V(P^{\prime}, S) > 0$
(the displaced fluid element has a lower volume per mass than other fluid in the
environment at $z+\xi$). Using $S^{\prime} - S \equiv \xi(dS/dz)$ and the 
relation $(\partial V / \partial S)_P = (T / C_P) (\partial V/\partial T)_P > 0$ 
we obtain the {\em stability condition} 
\begin{equation}   \label{eq-entr-grad}
\frac{dS}{dz} > 0 \, ,
\end{equation}
whence entropy has to increase with height for a convectively stable stratification.
For the temperature gradient this implies
\begin{equation}   \label{eq-entr-grad-v2}
\frac{dS}{dz} = \left( \frac{\partial S}{\partial T} \right)_P
                   \frac{dT}{dz}
                 + \left( \frac{\partial S}{\partial P} \right)_T
                   \frac{dP}{dz} > 0.
\end{equation}

To obtain Eq.~(\ref{eq-entr-grad-v2}) we have used assumption~2. It is at this point where inconsistencies come in 
once $\ell \gtrsim H_p$. The key criterion for stability in this analysis is the entropy gradient. Through the specific volume 
of the fluid element it can be linked to buoyancy. Using standard thermodynamical relations (see Chap.~9 of \cite{cox04b}) 
we obtain the stability criterion for the entropy gradient expressed in terms of the pressure and temperature gradient.

The corresponding criterion for chemically homogeneous stars reads:
{\em stars are convectively stable where their entropy increases as a function 
of radius}, i.e., if
\begin{equation}
  \frac{dS}{dr} > 0 \, .
\label{sstabil}
\end{equation}

For a perfect gas, $S \propto \ln[P T^{\gamma/(\gamma-1)}]$ whence $(\partial S/\partial T)_P = [\gamma / (\gamma-1)] T^{-1}$ 
and $(\partial S/\partial P)_T = P^{-1}$. With Eq.~(\ref{sstabil}) we thus find
\begin{equation}
  \frac{dS}{dr} = \frac{\gamma}{\gamma -1} \frac{d \ln T}{dr}
                + \frac{d \ln P}{dr} > 0 \, .
\label{konvcrit1}
\end{equation}
Using $dP/dr < 0$ (hydrostatic equilibrium), the chain rule of calculus, and 
the thermodynamic relation $\nabla_{\rm ad} \equiv (\gamma -1)/ \gamma$ the famous
{\em Schwarzschild criterion} for convective stability follows:
\begin{equation}
  \frac{d \ln T}{d \ln P} < \nabla_{\rm ad} \, .
\label{konvcrit}
\end{equation}
The spatial temperature gradient of a star expressed by $\nabla := (d\ln T/d\ln P)$
hence has to be smaller than the temperature change of a fluid element
due to adiabatic compression to ensure stable stratification with
respect to convection (see \cite{cox04b} for further details).

Eq.~(\ref{konvcrit}) is the standard form of the convective stability criterion used in astrophysics: for 
{\em unstable} stratification, $\nabla > \nabla_{\rm ad}$. In meteorology, the gradient of potential temperature 
is commonly used instead: $\nabla_{\rm \Theta} > 0$. Physically, these criteria are all equivalent.

\subsection{Conservation laws} \label{sec-2.5}

For convenience let us consider the case of radiation hydrodynamics for a one-component fluid
in the non-relativistic limit. Further extensions to this setup are the subject of other lectures described 
in this volume. The equations of hydrodynamics describe the time evolution of mass density $\rho$ 
(continuity equation),
\begin{equation}
    \partial_t\, \rho + {\rm div}\left( \rho \boldsymbol{u} \right) = 0,
\label{sec-2-eq_cont}
\end{equation}
and momentum density $\rho\boldsymbol{u}$ (compressible Navier-Stokes equation),
\begin{equation}
  \partial_t (\rho \boldsymbol{u})
 + {\rm div} \left( \rho (\boldsymbol{u} \otimes \boldsymbol{u}) \right) 
 = -{\rm div}\, \mbox{\boldmath$\Pi$} - \rho\, {\rm grad}\,\Phi,
\label{sec-2-eq_NSE}
\end{equation}
where $\partial f / \partial t$ is abbreviated by $\partial_{t} f$ and the fluid velocity is denoted by $\boldsymbol{u}$.
We first complete this set of equations before the meaning of all variables is explained just below.
The conservation law for the total energy density $\rho\,E$, where $E = (|\boldsymbol{u}|/2) + \varepsilon$, and thus
\begin{equation}
     \partial_t \left( \rho E \right)
  + {\rm div} \left( (\rho E + p) \boldsymbol{u} \right)
  = q_{\mathrm{source}}  + {\rm div} ( \mbox{\boldmath $\pi$} \boldsymbol{u} ) - \rho \boldsymbol{u}\cdot {\rm grad}\, \Phi,
\label{sec-2-eq_energy}
\end{equation}
is used to close this system. The pressure tensor $\mbox{\boldmath$\Pi$}$ in Eq.~(\ref{sec-2-eq_NSE}) is defined as 
\begin{equation} 
 \mbox{\boldmath$\Pi$} = p\, \mbox{\boldmath$I$} - \mbox{\boldmath$\pi$} \,, 
\label{sec-2-eq_pressure-tensor}
\end{equation} 
where $\mbox{\boldmath$I$}$ is the unit tensor with its components given
by the Kronecker symbol $\delta_{ik}$. The components of the tensor 
viscosity $\boldsymbol{\pi}$ in turn are given by 
\begin{equation}
 \pi_{ik} = \eta \left( 
            \partial_{x_k} u_i + \partial_{x_i}  u_k
          - \frac{2}{3} \delta_{ik} \, {\rm div}\, \boldsymbol{u} 
            \right) 
          + \zeta \delta_{ik} {\rm div} \, \boldsymbol{u} \,,
\label{sec-2-eq_visc-tensor}
\end{equation}
where $\partial f / \partial x_j \equiv\partial_{x_j} f$. From Eq.~(\ref{sec-2-eq_visc-tensor})
the standard form of the Navier-Stokes equation,
\begin{equation}
  \partial_t (\rho \boldsymbol{u})
 + {\rm div} \left( \rho (\boldsymbol{u} \otimes \boldsymbol{u}) \right) 
 = -{\rm grad}\, p + {\rm div}\, \boldsymbol{\pi} + \rho\, \boldsymbol{g} \,,
\label{sec-2-eq_NSE-std}
\end{equation}
is obtained and identifying the local sources sources and sinks of energy for the fluid,
\begin{equation}
   \partial_t  \left( \rho E \right)
 + {\rm div} \left( (\rho E + p)\, \boldsymbol{u} \right)
  = {\rm div} ( \boldsymbol{\pi} \boldsymbol{u} )
   - {\rm div}\, \boldsymbol{f}_{\rm rad} - {\rm div}\, \boldsymbol{h}
      + \rho \boldsymbol{u}\cdot \boldsymbol{g} + q_{\rm nuc} \,.
\label{sec-2-eq_energy-std}
\end{equation}

We now summarize the meaning of all variables not yet introduced. The dynamical 
variables ($\rho$, $\rho \boldsymbol{u}$, $\rho\, E$) are vector fields as function of
time $t$ and one, two, or three space coordinates $\boldsymbol{x}$.
Moreover, $e = e(\boldsymbol{x},t)$ is total energy density (sum of internal and kinetic one), 
while $E= e/\rho = E(\boldsymbol{x},t)$ is specific total energy or total energy per unit of mass,
and $\varepsilon = E - \frac{1}{2}|\boldsymbol{u}|^2 = \varepsilon(\boldsymbol{x},t)$ is the specific 
internal energy. The latter is related to the temperature $T=T(\rho,\varepsilon,\tilde{c})$ through 
an equation of state and we use $\tilde{c}$ to denote the different quantities specifying 
chemical composition. For idealized microphysics $T=T(\boldsymbol{x},t)$ is frequently used.
The gas pressure $p = p(\rho,T,\tilde{c})$ is also obtained from an equation of state and in Boussinesq 
approximation, $p=p(\boldsymbol{x},t)$. For zero bulk viscosity $\zeta$, the  viscous stress tensor 
{\boldmath$\pi$} is usually denoted as {\boldmath$\sigma$} and depends only on one microphysical
quantity, the shear or dynamical viscosity $\eta$, which follows from the kinematic viscosity: $\nu=\eta / \rho$.
Each of these are in general functions of $(\rho,T,\tilde{c})$. The radiative transport properties are
characterized by the radiative conductivity, $K_{\rm rad} = K_{\rm rad}(\rho,T,\tilde{c})$, which for idealized 
model systems is often a function of location, $K_{\rm rad} = K_{\rm rad}(\boldsymbol{x})$. It is related
to the radiative diffusivity $\chi_T = K/(c_p \rho) = \chi(\rho,T,\tilde{c})$ (knowing the specific heat 
at constant pressure) and requires the computation of opacity, $\kappa = \kappa(\rho,T,\tilde{c})$.
The equivalent of $K_{\rm rad}$ for the case of heat conduction is the conductivity $K_{\rm h}$.

Finally, $\boldsymbol{g} = -{\rm grad}\, \Phi$ is the solution of the Poisson equation 
${\rm div}\,{\rm grad}\, \Phi = 4\pi{\rm G}\rho$, where $G$ is the gravitational constant,
$g$ is the gravitational acceleration in vertical (or radial) direction, and $\boldsymbol{g}$ is
the vector of gravitational acceleration, $\boldsymbol{g}=(g,0,0)$, if the vertical component
is given first. In practice, the gravitational force is often computed from (approximate) 
analytical solutions or held constant. 
Since in all cases of interest here $q_{\rm nuc}$ is a function of local thermodynamic 
parameters ($\rho$, $T$, chemical composition, cf.\ \cite{kippenhahn94b}), we find that 
Eqs.~(\ref{sec-2-eq_cont}), (\ref{sec-2-eq_NSE-std}) and (\ref{sec-2-eq_energy-std}) 
together with equations for the radiative and conductive heat flux, $\boldsymbol{f}_{\rm rad}$ 
and ${\rm div}\, \boldsymbol{h}$, as well as (\ref{sec-2-eq_visc-tensor}) form a closed system 
of equations provided the material functions for $\kappa$, $K_{\rm h}$, $\nu$, $\zeta$, and 
the equation of state are known, too. Usually, they are used in stellar modelling in pretabulated form.

At high densities, heat conduction contributes: $q_{\rm cond} = -{\rm div} F_{\rm cond} = K_h \nabla T$. 
Otherwise, the non-convective energy flux is mostly due to radiation, $q_{\rm rad} = -{\rm div} F_{\rm rad}$,
for which inside stars the diffusion approximation $F_{\rm rad} = -K_{\rm rad}\nabla T$ holds.
In the photosphere, the radiative transfer equation is solved, along several rays, usually assuming 
also LTE, and grey or binned opacities. Its stationary limit reads (see \cite{mihalas84b} who also
provides validity ranges for this form)
\begin{equation}
{\bf r}\cdot \nabla I_{\nu} = \rho\,\kappa_{\nu} (S_{\nu}-I_{\nu}),
\end{equation}
where ${\bf r}$ denotes position, $I_{\nu}$ intensity at frequency $\nu$, $\kappa_{\nu}$ opacity for given $\nu$, 
and $S_{\nu}$ is the source function which in LTE equals the Planck function $B_{\nu}$. Angular and frequency 
integration of intensity then yield $q_{\rm rad}$ and $F_{\rm rad}$.

In numerical simulation of stellar convection the hydrodynamical equations are solved
for a finite set of grid cells, with special geometrical assumptions on the simulation box:
for whole stars or parts thereof, spheres and spherical shells are common which implies using
a polar geometry and an appropriate coordinate system. Box-in-a-star (``local simulations'' of
solar granulation and the like) and ``star-in-a-box'' concepts are also popular, since they can rely
on Cartesian geometry. In more recent simulation codes, mapped grids, Ying-Yang grids,
and grid interpolation in principle allow for an arbitrary or ``mixed geometry'' (transition from
a Cartesian grid near the centre to a polar one near the surface of a star). Boundary conditions
for those simulations are either idealized (such as setting the vertical velocity zero for
some layer), or are based on realistic physical models, or lateral periodicity is assumed
(for global, whole sphere simulations this is trivial, for local simulations in Cartesian geometry
it is often a good approximation).

\subsection{The challenge of scales} \label{sec-2.6}

The huge range of temporal and spatial scales is one of the main challenges in modelling 
turbulent convection in astro- and geophysics. Such flows are characterized  by very 
large scales $L$ on which stratification, heating, and cooling act.
Viscous dissipation in turn occurs on very small scales $l_d$. The very high temperatures 
of astrophysical fluids makes them distinct in comparison with geophysical flows due to the 
very large mean free path for photons. As a result, in astrophysics the Prandtl number is
computed from the ratio of kinematic molecular viscosity to the radiative diffusivity instead of molecular 
heat diffusivity and consequently, $\rm Pr \ll 1$ instead of ${\rm Pr} \sim O(1)$. For the Sun, $L$ may be 
taken as large as $180000$~km compared to $l_d$ which for the interior is in the range of 1~cm to 10~cm. 
As a result, the Reynolds number $\rm Re = U(L)\,L / \nu$ for the convective zone is of order 
$10^{10}$ to $10^{14}$ for typical velocities $U(L)$ depending on the specific length scales $L$ considered whereas 
$\rm Pr$ is in the range of $10^{-6}$ to $10^{-10}$. In comparison, the convective boundary of the atmosphere 
of the Earth has $L \sim 1\,\rm km$ with $l_d \sim 1\,\rm mm$ and $\rm Re \sim 10^8$ at $ \rm Pr \sim 0.7$.
For oceans on Earth $L$ may range form a few km to a few $10^3\,\rm km$ with $l_d \sim 1\,\rm mm$ 
and $\rm Re$ in the range of $10^9$ to $10^{12}$ whereas Pr is around 6 to 7 (varying by a factor of 2
according to specific local conditions). Hence, the scale range for these systems is very similar.
Differences among these systems originate from the Prandtl number, the boundary conditions, the 
specific microphysics with phase transitions, and, in the solar or stellar case, of course the possibility 
of magnetic fields interacting with the flow.

For simulations of solar convection one can make more specific estimates on Reynolds and Prandtl
numbers, P\'eclet numbers, and required grid sizes. A detailed calculation can be found, for instance,
in Sect.~2.2 of \cite{kupka17b}. This issue is addressed again in Sect.~\ref{sec-4} and~\ref{sec-5} below.

\newpage

\section{Modelling of convection}  \label{sec-3}

Most convection models which are currently used in astrophysical calculations assume
local isotropy and horizontal homogeneity of the flow as well as that non-local transport 
can be described as a diffusion process. Geometrical properties of the flow are generally
neglected. Some consider a range of bubbles or ``eddy-sizes'' and dispersion lengths which 
all depend on local quantities such as temperature or pressure. None of these assumptions
strictly hold for convection in astrophysics! Where does this preference for such simplifications
come from and what are its consequences?

\begin{figure}[h]
\centering
\includegraphics[width=\textwidth,clip]{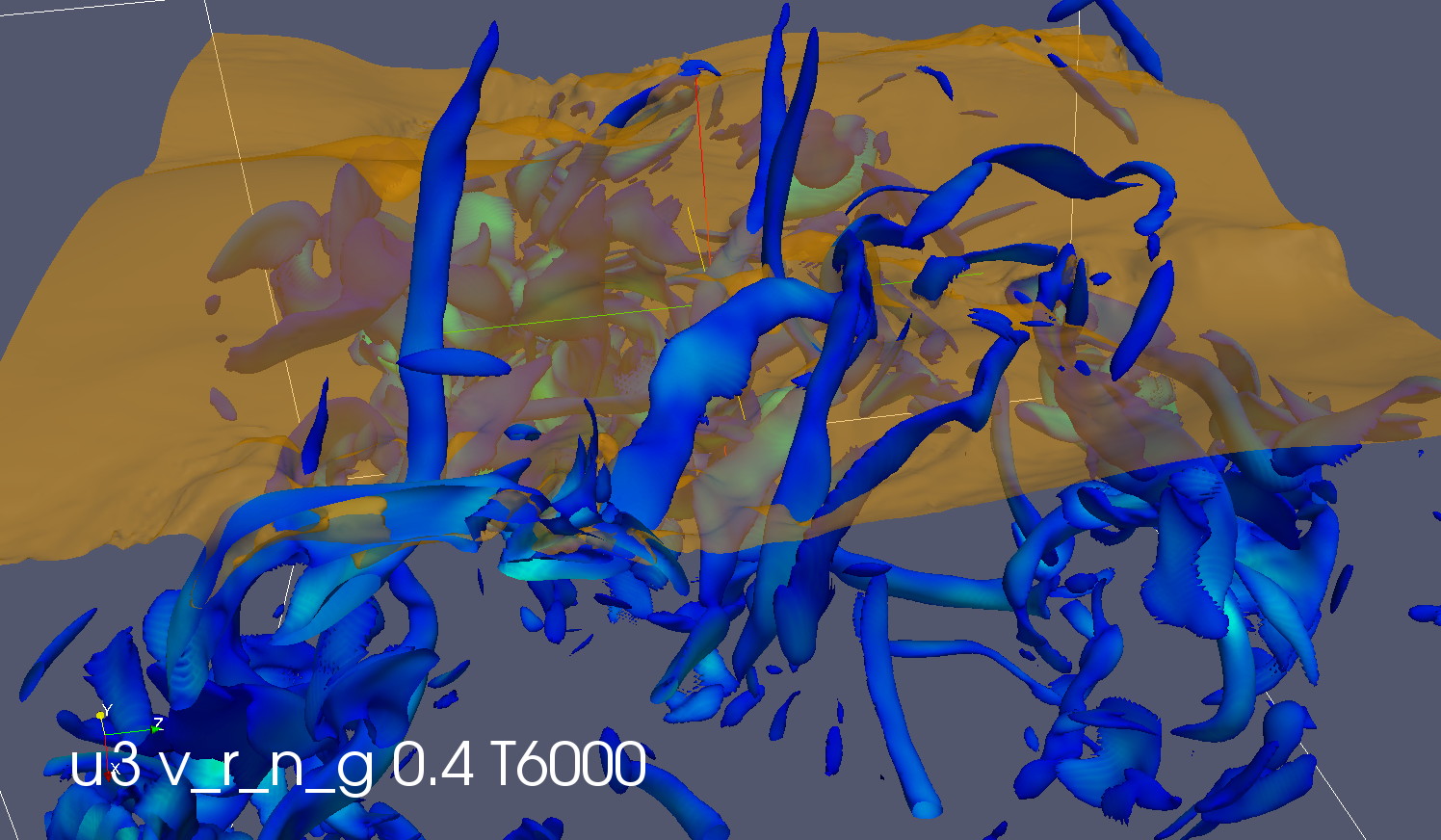}
\caption{Numerical simulation of convection at the solar surface, presented originally in \cite{muthsam11b}.
             The isosurface where $T = 6000\,\rm K$ is shown by a dark yellow shade. Blue tubes mark
             where a certain value of the magnitude of vorticity is reached. The domain shown is a region with
             double grid refinement, 1.2~Mm wide horizontally and 1~Mm deep, at a resolution of 3.7~km horizontally
             and 2.5~km vertically. It is embedded inside a domain 2.5~Mm to 2.8~Mm wide and 1.9~Mm deep at
             a grid spacing of 5.1~km vertically and 7.4~km horizontally. The latter is the resolution at which
             the structures visible at this instant of time have formed. This outer box resides inside one which
             is 3.7~Mm deep and 6~Mm wide and which has half of its vertical and one third of its horizontal resolution
             (figure by courtesy of H.J.~Muthsam).}
\label{fk.fig-1}       
\end{figure}

\subsection{MLT and phenomenological models} \label{sec-3.1}

If we consider a numerical simulation of convection at the surface of a star such as our Sun 
(e.g., as in \cite{stein00b,freytag12b,muthsam10b,robinson03b} or in the comparison of 
\cite{beeck12b}), it is immediately evident that the flow field is non-isotropic and inhomogeneous
(there are structures of up- and downflows), a whole range of flow structures co-exist, and it is not at all 
evident why such a flow could be successfully described by a diffusion process (see Fig.~\ref{fk.fig-1}). 
This holds especially since the background quantity driving the flow, the entropy gradient, varies substantially 
on the scale of the flow structures visible at the surface while it also leads to an inhomogeneous pattern in emitted 
intensity (solar granulation) that coincides well with observations (cf.\ \cite{stein00b}).

\begin{figure}[h]
\centering
\includegraphics[width=\textwidth,clip]{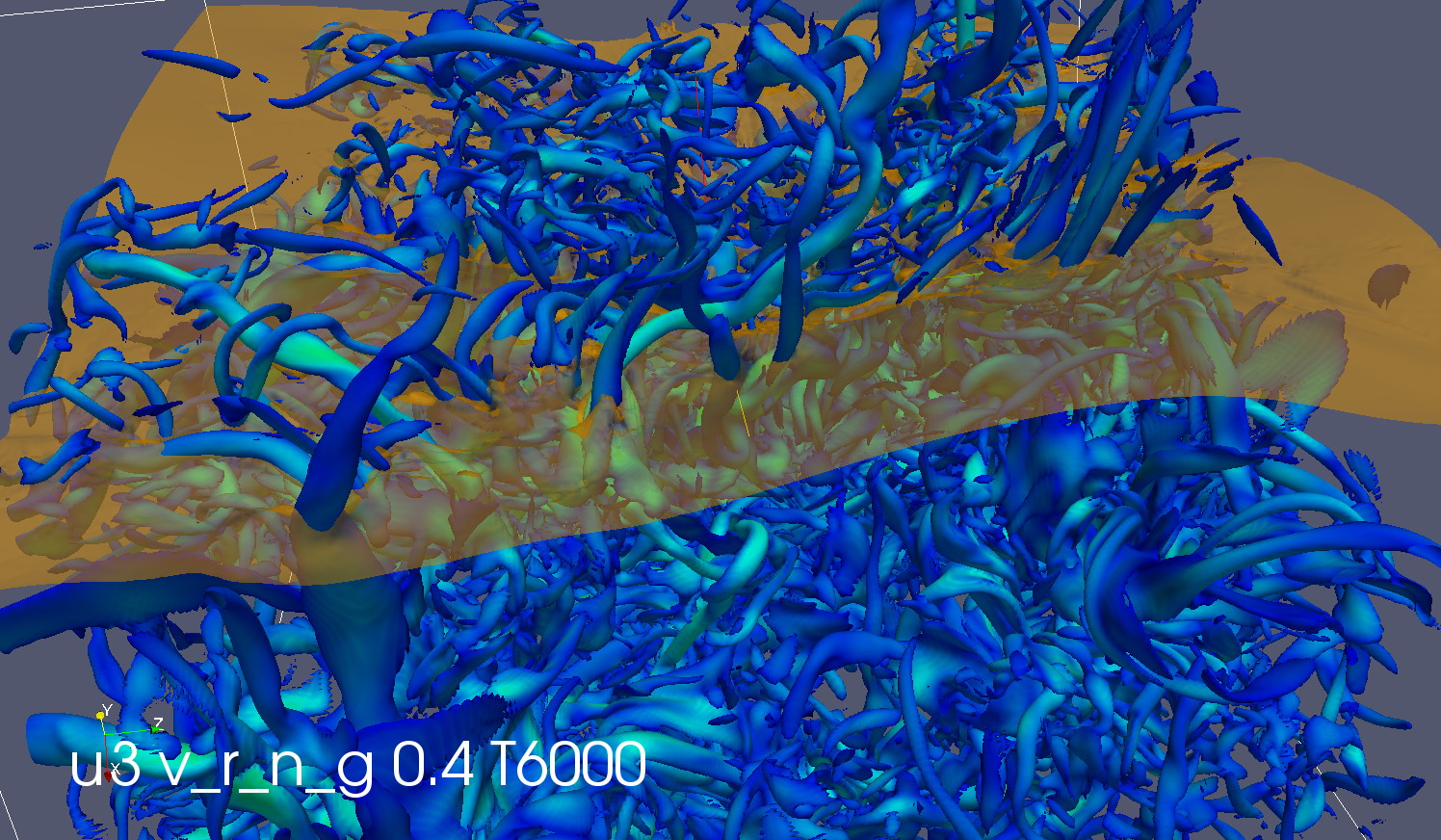}
\caption{Numerical simulation of convection at the solar surface, presented originally in \cite{muthsam11b}.
             This is the same simulation as shown in Fig.~\ref{fk.fig-1}, but after further evolution of about 143~s.
             In this time shearing stresses accounted for on the fine computational grid used for
             this simulation have led to a highly turbulent flow (figure by courtesy of H.J.~Muthsam).}
\label{fk.fig-2}       
\end{figure}

\subsubsection{The concept of turbulent viscosity} \label{sec-3.1.1}

Even for present supercomputers it is challenging to fully reveal the turbulent nature of stellar 
convection (see Fig.~\ref{fk.fig-2}). The ideas traditionally used in convection models of course
are much older. In his ``Essai sur la th\'eorie des eaux courantes'' J.~Boussinesq introduced 
the concept of turbulent viscosity \cite{boussinesq1877b}. Contrary to molecular viscosity, its computation 
is model dependent, since it is not a property of a {\em fluid}, but a property of the {\em flow}.
Its underlying idea is to assume that the main effect of turbulence is a boost of molecular 
viscosity. Thus, we express velocity fluctuations relative to this mean as
$u_i^{\prime} = u_i - <u_i> = u_i - U_i$ for a velocity field $u_i$ (written in index notation)
that has an average of $U_i$. The Reynolds stress is defined to be $<u_i^{\prime} u_j^{\prime}>$
and is assumed to be proportional to the mean strain rate $S_{ij} = (U_{i,j}+U_{j,i})/2$ where
$f_{,j} = \partial f / \partial x_j$. The proportionality constant in this relation is the 
{\em turbulent viscosity}, $\nu_{\rm turb}$, which has to be computed from a model and allows 
the introduction of an effective or dynamical viscosity, $\nu_{\rm eff} = \nu + \nu_{\rm turb}$,
which is used to replace $\nu$ in expressions for the (mean) strain rate. In the end this model assumes
that in a turbulent flow the main effect of small (and in computations possibly unresolved) scales
on large scales is that of an ``extra viscosity''. 

The first and most simple model for $\nu_{\rm turb}$ had been to take it constant. But
this assumption already fails for turbulent boundary layers where $\nu_{\rm turb}$
must vary along the cross-flow direction to obtain sensible results (cf.\ \cite{pope00b}).
To solve this problem L.~Prandtl proposed in 1925 his mixing length theory \cite{prandtl25b}.
It assumes $\nu_{\rm turb} \sim u_{\rm rms}\ell$, with $u_{\rm rms} \sim \ell S$ and 
$S=(2\,S_{ij}\,S_{ij})^{1/2}$. The {\em mixing length} $\ell$ is taken from the geometry of the 
system or other constraints. In particular for pipe flow or boundary layer flow, $\ell$ is taken 
as the distance to the (nearest) boundary. For turbulent boundary layer flow this model 
can explain some basic data (\cite{pope00b}) and is clearly an improvement 
in comparison to the model where $\nu_{\rm turb} = \rm constant$.

\subsubsection{Turbulent diffusivity and the MLT model of stellar convection} \label{sec-3.1.2}

Similar to turbulent viscosity the concept of a {\em turbulent diffusivity} ``generalizes'' the law of 
gradient diffusion which had been proposed to model the transport of heat and concentration
on molecular scales. Originally introduced in 1915 by G.I.~Taylor \cite{taylor1915b} and
further developed in 1929 by D.~Brunt \cite{brunt29b}, the idea behind turbulent diffusivity
is to consider a driving gradient which would lead to a transport of some quantity at
the molecular level (heat conduction or, alternatively, radiation). A gradient diffusivity 
$\chi$ can hence be associated with a (diffusive) flux $-\chi \nabla <\phi>$. The idea
now is to replace $\chi$ by $\chi_{\rm eff} = \chi + \chi_{\rm turb}$ and approximate
the (advective) transport of $\phi$ through a turbulent flow with a {\em gradient diffusion 
hypothesis}: $\mbox{$<u'_i\phi^{\prime}>$} \sim -\chi_{\rm turb} \nabla <\phi>$. If $\chi$ is the heat 
diffusivity, related to the kinematic viscosity $\nu$ by the Prandtl number $\rm Pr = \nu / \chi$, 
we obtain an approximation for the turbulent heat flux.

Together with the concept of turbulent viscosity this provides the basis for stellar mixing length theory 
(MLT) as introduced in 1932 by L.~Biermann \cite{biermann32r}. Just as in linear stability analysis 
the driving gradient can either be given by entropy ($dS/dr$), potential temperature ($\nabla_{\Theta}$), or 
superadiabaticity ($\nabla - \nabla_{\rm ad}$), the latter being most popular in astrophysics.
In the end stellar MLT is a (down-) gradient diffusion model for the heat flux with a particular 
description for deriving the turbulent heat conductivity. Various derivations of it yield equivalent 
results (cf.\ \cite{bv58b,canuto96c,cox04b}). The catch of this approach is that it is completely 
unclear whether it is applicable beyond the linear regime considered during its derivation.

In \cite{canuto96c} a particularly compact variant of the MLT is discussed which
is repeated here for convenience. The convective flux is computed from

\begin{equation}  \label{eq_local_Fc_formula}
   F_{\rm conv} = K_{\rm turb} \beta = 
   K_{\rm rad} T H_p^{-1} (\nabla-\nabla_{\rm ad})\Phi(\nabla-\nabla_{\rm ad},S^{*})
\end{equation}
where $\nabla > \nabla_{\rm ad}$ with $\nabla=\partial \ln T / \partial \ln P$ and $\nabla_{\rm ad}=(\partial \ln T / \partial \ln P)_{\rm ad}$.
Here, $K_{\rm rad} = c_p\rho\chi$ is the radiative conductivity (in the diffusion approximation of radiative transfer),
$c_p$ is the specific heat at constant pressure, $\rho$ is the density, $T$ is the temperature, $P$ is the pressure
(usually the sum of gas and isotropic radiative pressure), and $H_p = -P/(\partial P/\partial r)$ is the local pressure scale 
height along the radius coordinate $r$. The turbulent (heat) conductivity $K_{\rm turb}$ has to be computed from a model
(the function $\Phi$) while the superadiabatic temperature gradient is calculated from basic quantities of stellar structure:
\begin{equation}
   \beta = -\left(\frac{dT}{dr}-\left(\frac{dT}{dr}\right)_{\rm ad}\right) = 
          T H_p^{-1} \left(\nabla-\nabla_{\rm ad}\right).
\end{equation}
The convective efficiency is expressed by the squared ratio of the (radiative) diffusion time scale 
to the buoyancy time scale, which we call $S^{*}$ here to avoid its confusion with entropy. It can be computed
from
\begin{equation}
    S^{*} = {\rm Ra}\cdot{\rm Pr}
      = \frac{g\alpha_{\rm v}\beta l^4}{\nu\chi}\cdot\frac{\nu}{\chi}\,,
\end{equation}
where $\alpha_{\rm v}$ is the volume expansion coefficient, $g$ the (radial component of)
gravitational acceleration, and $l$ is the {\em mixing length}. This allows a compact expression
for the turbulent heat conductivity of MLT: $\Phi(S^{*})$ is given by $\Phi(S^{*}) = \Phi^{\rm MLT}$ with
\begin{equation} \label{eq_PhiMLT}
   \Phi^{\rm MLT} = \frac{729}{16} (S^{*})^{-1} \left((1 + \frac{2}{81} S^{*})^{1/2}-1\right)^3,
\end{equation}
and where $S^{*}$ is computed from
\begin{equation}
  S^{*} = \frac{81}{2} \Sigma, \quad \Sigma = 4 A^2 (\nabla-\nabla_{\rm ad}), \quad
    A = \frac{Q^{1/2} c_p \rho^2 \kappa l^2}{12 a c T^3} \sqrt{\frac{g}{2 H_p}}\,,
\end{equation}
and $Q = T V^{-1}(\partial V/\partial T)_p = 1-(\partial \ln \mu/ \partial \ln T)_p$ is
the variable, average molecular weight.
Finally, the {\em mixing length} $l$ is usually parametrized as
\begin{equation}   \label{eq_mixinglength}
   l = \alpha H_p \,,
\end{equation}
where $\alpha$ is calibrated by comparison with some data or another, integral constraint.

\subsubsection{Countergradient flux and non-locality, shortcomings of popular models}

Already in 1947, C.H.B.~Priestley and W.C.~Swinbank realized \cite{priestley47b} that
the downgradient diffusion model is insufficient to explain observed convective turbulence in the 
planetary boundary layer of the atmosphere of the Earth (note that this was more than 10 years
before publication of the variant of MLT which is commonly used now in astrophysics, \cite{bv58b}).
The problem they discovered is that the Schwarzschild criterion predicts a convective instability to 
exist only where $\nabla_{\Theta} > 0$ (or $\nabla > \nabla_{\rm ad}$), an assumption intrinsic to 
the MLT prescription for the computation of the convective heat flux. However, turbulence 
is observed in the atmosphere of the Earth next to a convectively unstable zone also
where $\nabla_{\Theta} < 0$. This points towards a non-local nature of the flow in the 
planetary boundary layer. As is discussed below (Sect.~\ref{sec-5.1}), regions for
which $F_{\rm conv} \neq 0$ while $\nabla < \nabla_{\rm ad}$ are called overshooting zones
in astrophysics. A first modification of the downgradient model was proposed in \cite{priestley47b}.
Deardorff \cite{deardorff66b,deardorff72b} then demonstrated the essential role of non-local fluxes
both from the viewpoint of observational data and from theoretical considerations and introduced the 
concept of countergradient flux into models of the convective planetary boundary layer of the Earth. In 
his model, $F_{\rm conv} > 0$ is compatible with $\nabla_{\Theta} < 0$ ($\nabla < \nabla_{\rm ad}$).
We return to this question in Sect.~\ref{sec-5.1}.

The main shortcomings of the most popular models of convection used in astrophysics
can be summarized as follows:
\begin{itemize}
\item the commonly used models are of pure diffusion type with a locally specified diffusivity;
\item they ignore non-local production and thus cannot explain the mixing next to but outside 
        a convectively unstable region;
\item they ignore apparent geometrical properties such as spatial inhomogeneity of a flow;
\item for use in astrophysics they require a calibration with solar data, but global quantities 
        such as luminosity are used for this purpose despite the quantity calibrated (primarily the mixing length) 
        describes effects at a more local scale and has been demonstrated to vary over the 
        Hertzsprung-Russell diagram (HRD) and as a function of radius.
\end{itemize}        
Alternatives to these models have resulted in the demand for numerical simulations needed for
stars all over the HRD. But what about less expensive non-local models of convection?

\subsection{Ensemble and volume averages} \label{sec-3.2}

Recalling the huge range of length (and time) scales (see Sect.~\ref{sec-2.6}) on 
which convection in stars operates some kind of averaging has to be applied 
to the dynamical equations introduced in Sect.~\ref{sec-2.5}. Two different types
of averages are useful in this context. The first one is a {\em volume averaged interpretation}
of all functions $f(t,x,y,z)$ that appear in the dynamical equations. The goal here is to 
explicitly account for the dynamically most important length scales in the modelling. 
In the {\em Large Eddy Simulation (LES)} approach these are the length scales 
where most of the kinetic energy is concentrated, such as the up- and downflow
patterns of stellar granulation. This allows for numerical simulations with realistic microphysics.
The second approach is based on an {\em ensemble average interpretation} of $f(t,x,y,z)$ and
allows the computation of statistical properties of the flow as expressed by the average $<f(t,x,y,z)>$.
This is the basis of (semi-) analytical convection models and typical quantities computed
with this approach are $F_{\rm conv}$ (the convective or enthalpy flux), $P_{\rm turb}$ 
(the turbulent pressure), or $v_{\rm rms}$, the root mean squared (flow) velocity.

Let us return to the first type of average. In a grid based numerical approach to solve
the hydrodynamical equations the volume average interpretation of $\rho(t,x)$, $\rho u(t,x)$, 
and $\rho e(t,x)$ occurs quite naturally: either the function values given at the grid points
are considered to be interpolation nodes for a numerical approximation which has to
ensure to conserve the basic variables within a grid cell as in a conservative finite difference 
scheme. Or the function values on the grid are considered as numerical approximations
to the averages of the basic variables over the grid cell (finite volume approach, conservative
by construction). Both concepts to solve the hydrodynamical equations thus are quite
naturally compatible with the idea of a volume average. The LES approach in a strict sense
is closely related to the finite volume approach, but not identical with it. As indicated the
idea is to choose a grid such that scales carrying most of the kinetic energy are explicitly
resolved by the grid. For simulations of stellar granulation this includes those
spatial scales on which radiative cooling takes place at the stellar surface. Smaller scales 
are accounted for through a ``subgrid scale model'' or simply, in the sense of the concept
of turbulent viscosity discussed in Sect.~\ref{sec-3.1.1}, by some form of extra 
(artificial, etc.) viscosity. The influence of scales larger than the domain size has to be
taken care of by boundary conditions. As starting point for such a numerical simulation 
one assumes ``typical'' initial conditions and first simulates the (kinetic and thermal) relaxation
of the system before a statistical interpretation of the results based on a quasi-ergodic hypothesis
is done (cf.\ \cite{kupka17b}). Continued time integration is used to generate ensembles
and ``typical cases'' where the latter may be inspected by some visualization software.
This approach is based on the assumption that a sufficiently long time integration provides statistical
realizations of the physical system with the same probability distribution as if they
were generated directly from a whole set of simulations with (slightly) different, randomly generated 
initial conditions. The computation of averages over long time intervals allows a comparison of the 
simulation with astronomical observations and direct ensemble averages. But this approach is of course
computationally quite expensive (see again \cite{kupka17b}).

The goal of {\em ensemble averaged models} on the other hand is to directly compute functions such as
$<\rho(t,x,y,z)>$ where the ensemble consists of different realizations of $\rho$ (obtained from
``sensible'' initial conditions which differ only by a small amount from each other). Of course,
these have to be compatible with the boundary conditions. There are clear differences for such
models in comparison with the strict derivation of the Navier-Stokes equations (NSE). The 
macro-states considered for a turbulent flow are much more complex than the ``fluid-elements'' 
invoked in the classical, phenomenological models such as MLT (Sect.~\ref{sec-3.1.2}). There
is no generally valid and fully self-consistent procedure known to derive such models from first 
principles, i.e., the NSE, only. Thus, convection models, just as any other models of turbulent flows, 
need additional hypotheses, the {\em closure approximations}. These might be simply inspired
from analyses of laboratory systems, but the non-linearity of this problem and its sensitivity
to boundary conditions make an approach based just on data from laboratory scenarios highly
incomplete and likely to fail in applications to astrophysical flows. Hence, a more general strategy 
is needed.

\subsection{Reynolds stress approach} \label{sec-3.3}

An ensemble average can be computed straight from a numerical simulation
assuming that the quasi-ergodic hypothesis holds for it. When deriving (semi-) 
analytical models it is common to first perform a splitting of the fields of
the dynamical variables into an often slowly varying background and 
any fluctuations around it. The background state is simply taken to be
the ensemble average of the basic variable, say $\overline{\rho}$.
This {\em Reynolds decomposition} or {\em Reynolds splitting} was first 
suggested in 1894 by \cite{reynolds1894b}. Formally, an equation for $A(t,x,y,z)$ 
is additively split by introducing $A = \overline{A} + A'$. The dynamical equation for $A$ is 
averaged to obtain an equation for $\overline{A}$. Since the NSE are non-linear, the equation for
$\overline{A}$ actually depends on averages over products, $\overline{A'A'}$, of the fluctuations
around the mean state. 

This gives rise to the idea of a moment expansion of the hydrodynamic equations,
which was first suggested in 1925 by Keller and Friedmann, \cite{keller25b}. This procedure
requires to subtract the dynamical equation for $\overline{A}$ from the complete equation 
for $A$ to obtain a dynamical equation for $A'$. A dynamical equation for $\overline{A'A'}$ 
is obtained from computing the product of the latter, averaging it, and using basic calculus and algebra. 
But due to the non-linearity of the underlying equations an infinite hierarchy of moments is obtained 
this way, the basic closure problem of hydrodynamics. To proceed from there requires 
assumptions in addition to the basic equations themselves. This is its main disadvantage 
over the volume average. The additional assumptions for closing ensemble averaged 
moment equations directly involve the physics of the large, energy carrying scales 
and those are very difficult to model analytically.

One-point closure turbulence models are the most widely used class of such models.
Their main idea is to perform the ensemble averaging directly in physical space $(x,y,z)$.
At lowest order this generates equations for the mean thermal structure, i.e.,
temperature $\overline{T}$ and pressure $\overline{P}$, as well as for a non-zero mean flow.
Fluctuations around these mean values describe the ``turbulent component'' of the flow:
$w = W - \overline{W}$, $\theta = T - \overline{T}$. The dynamical equations for those are
used to derive dynamical equations for the second order moments (SOMs) of the basic
variables: the quantities $\overline{w^2}$ and $\overline{\theta^2}$ quantify the kinetic
and thermal (potential) energy contained in the turbulent component of the dynamical fields
and their cross correlation $\overline{w\theta}$ is associated with the contribution of
the turbulent component to the mean heat flux in the system. The next higher order
statistics, the third order moments (TOMs) describe non-local transport by the 
turbulent component (flux of turbulent kinetic energy) and basic asymmetries of the
flow field (skewness).

Usually though not necessarily, the ensemble averaged equations are constructed
for the horizontal average of the fields. Various isotropy assumptions are common as well. 
The most popular  closure assumptions are based on expressing all higher order moments 
in terms of SOMs and sometimes also TOMs. For instance, the fourth order moments 
are frequently assumed to follow a Gaussian distribution. If no further restrictions
are put on the TOMs in this case, this approach is known as quasi-normal (QN) 
approximation. ``Local models'' on the other hand assume that the TOMs are zero
contrary to ``non-local models'' which have non-zero TOMs. Instead of the plain QN 
approximation the so-called damped QN approximation is often preferred which
limits the size of the TOMs (cf.\  \cite{andre76b,andre76c}) by explicit clipping
or by damping through boosting some of the closure terms.

An attractive property of this approach is that it is straightforward to understand the physical 
meaning of at least the lower order moments. The mean structure of the object modelled is 
given by the averages of $\overline{T}$ and $\overline{P}$. The second order moments describe 
the effects of convection on the mean structure: the enthalpy flux $F_{\rm conv} = c_p \rho \overline{w\theta}$
can modify the thermal structure in comparison with pure radiative or conductive heat transfer.
The {\em turbulent pressure} $P_{\rm turb} = \rho \overline{w^2}$ can change the hydrostatic
equilibrium structure. In addition, if the mean structure is perturbed by a large scale velocity field 
(global oscillations), the turbulent fields provide feedback and interact (see \cite{houdek15b}
for further details).  Third order moments describe non-local transport through advection and
are thus related to quantities such as the filling factor (fraction of horizontal area covered by upflows,
for example) and, as already indicated, the skewness of the velocity and temperature field.
Standard local models of convection such as mixing-length theory consider horizontal averages 
of the mean structure and the SOMs {\em and} ignore non-local transport: the TOMs are set to zero. 
The SOMs are treated using algebraic relations and concepts from the diffusion approximation are used

Non-local convection models thus have to first and foremost suggest a non-trivial approximation for 
the TOMs. An advanced model of this type was derived in \cite{canuto92b,canuto93b} and \cite{canuto98b}. 
It consists of 5 differential equations which describe the time evolution of turbulent kinetic energy 
$\overline{q^2} = \overline{u^2} + \overline{v^2}  + \overline{w^2}$  (i.e.\ the sum over squared fluctuations 
of horizontal ($u$, $v$) and vertical ($w$) velocity components), of temperature fluctuations 
$\overline{\theta^2}$, of the (convective) temperature flux $\overline{w\theta}$, and of the vertical component 
of turbulent kinetic energy, $\overline{w^2}$. An additional differential equation provides
the kinetic energy dissipation rate $\epsilon$. Heat loss effects on temperature fluctuations are
accounted for by algebraic closures derived from turbulence modelling (see also \cite{kupka02b}
for further details). Compressibility effects and anisotropy effects are included in
this model following \cite{canuto93b}, see also \cite{kupka99b}. Various models for the TOMs 
have been suggested in this framework, \cite{canuto92b,canuto93b,canuto98b,canuto01b}.
These have been applied, among others, to models of the convective planetary boundary layer of the 
Earth (see \cite{canuto94b,canuto01b}). That work included comparisons to numerical simulations and
to the Deardorff and Willis laboratory experiment \cite{deardorff85b}.

To compare the complexity of this approach with MLT a variant of the Reynolds stress convection 
model of \cite{canuto92b,canuto93b,canuto98b} as used in \cite{kupka99b} is given in the following. 
Using the notation of \cite{canuto98b} and abbreviating $\partial_t \equiv \partial/\partial t$
and $\partial_z \equiv \partial/\partial z$ this convection model reads as follows:
\begin{equation}  \label{eq_TKE}
  \partial_t K + D_{\rm f}(K) = g \alpha_{\rm v} J - \epsilon
        + \frac{1}{2}C_{ii} + \partial_z (\nu \partial_z K), 
\end{equation}
\begin{equation}
 \partial_t(\frac{1}{2}\overline{\theta^2})
      + D_{\rm f}(\frac{1}{2}\overline{\theta^2})  =  \beta J
        - \tau_{\theta}^{-1}\overline{\theta^2} 
        + \frac{1}{2}\partial_z (\chi
          \partial_z \overline{\theta^2}) + \frac{1}{2}C_{\theta},
\end{equation}
\begin{equation}
  \partial_t J + D_{\rm f}(J)  =  \beta \overline{w^2}
        + \frac{2}{3} g \alpha_{\rm v} \overline{\theta^2}  
        - \tau_{{\rm p}\theta}^{-1} J
        + \frac{1}{2}\partial_z (\chi \partial_z J)    
        + C_3 + \frac{1}{2}\partial_z (\nu \partial_z J),
\end{equation}
\begin{equation}
    \partial_t(\frac{1}{2}\overline{w^2}) + D_{\rm f}(\frac{1}{2}
        \overline{w^2}) = - \tau_{\rm pv}^{-1}(\overline{w^2}-\frac{2}{3}K) 
        + \frac{2}{3} g \alpha_{\rm v} J - \frac{1}{3}\epsilon 
        + \frac{1}{2}C_{33} + \frac{1}{2}\partial_z (\nu \partial_z \overline{w^2}),
\end{equation}
\begin{equation}  \label{eq_epsilon}
   \partial_t \epsilon + D_{\rm f}(\epsilon) =
     c_1 \epsilon K^{-1} g \alpha_{\rm v} J - c_2 \epsilon^2 K^{-1}
       + c_3 \epsilon \tilde{N} 
       + \partial_z (\nu\partial_z \epsilon),
         \quad \tilde{N} \equiv \sqrt{g \alpha_{\rm v} |\beta|},
\end{equation}
\begin{equation}    \label{eq_diffeps}
   D_{\rm f}(\epsilon) \equiv \partial_z(\overline{\epsilon w})
   = -\frac{1}{2}\partial_z
   \left[(\nu_{\rm t} + \chi_{\rm t})\partial_z \epsilon \right].
\end{equation}
To calculate the mean stratification one has to solve these equations alongside
\begin{equation}   \label{eq_hydrostat}
   \partial_z (P + p_{\rm t})  =  - g \rho, 
\end{equation}
\begin{equation}  \label{eq_Temp}
   c_{\rm v}\rho \partial_t T +
            \rho \partial_t K = -\partial_z
            (F_{\rm r} + F_{\rm c} + F_{\rm k}),
\end{equation}
which are the equations of hydrostatic equilibrium and of flux conservation. They have
to be extended by an equation for mass conservation and for radiative transfer.
Non-locality is represented by the terms $D_{\rm f}(K)= \partial_z (\overline{q^2\,w}/2)$, 
$D_{\rm f}(\frac{1}{2}\overline{\theta^2}) = \partial_z(\overline{w\,\theta^2}/2)$, 
$D_{\rm f}(J) = \partial_z(\overline{w^2\,\theta})$, and 
$D_{\rm f}(\frac{1}{2}\overline{w^2})=\partial_z (\overline{w^3}/2)$, which require closure 
approximations for the third order moments $\overline{q^2\,w}$, $\overline{w\,\theta^2}$, 
$\overline{w^2\,\theta}$, and $\overline{w^3}$. An example for the latter is the
downgradient model Eq.~(\ref{eq_diffeps}) for $\overline{\epsilon w}$. Instead of closing
with only lower order moments, also other moments of the same order can be used. Often
though not necessarily these provide better performance: 
$\overline{\epsilon w} \approx 0.6\, \overline{q^2\,w} / \tau$ with $\tau = \overline{q^2} / \epsilon$
has been found in \cite{kupka07e} to be a much more accurate closure than Eq.~(\ref{eq_diffeps}).

The other variables introduced in Eq.~(\ref{eq_TKE})--(\ref{eq_Temp})
are the following ones: $K=\overline{q^2}/2$ is the turbulent kinetic energy
(per mass unit). $\overline{\theta^2}/2$ are the squared fluctuations of the temperature field 
around its mean value and $J=\overline{w\,\theta}$ is the cross-correlation between vertical
velocity and temperature, both introduced further above. Moreover, $\alpha_{\rm v}$ is the volume 
compressibility (for a perfect gas this is $=1/T$) and $\beta$ is the superadiabatic gradient.
The  radiative (thermal) diffusivity $\chi = K_{\rm rad}/({c_p \rho})$ and the kinematic viscosity 
are related to each other through $\nu = \chi {\rm Pr}$. The variables denoted
with the letter $\tau$ are all time scales. Equations~(25a), (27b), and (28b) of \cite{canuto98b}
can be used to relate these to the dissipation time scale $\tau = 2 K / \epsilon$. 
Compressibility effects are included through the additive terms $C_{ii}, C_{\theta}, C_3, 
C_{33}$ taken from Equations~(42)--(48) of \cite{canuto93b}. Depending on their
actual size some of their contributions can also be neglected, see \cite{kupka99b}. 
In Eq.~(\ref{eq_diffeps}) the turbulent viscosity $\nu_{\rm t} = C_{\mu} K^2/\epsilon$
is introduced where $C_{\mu}$ is a constant given by Eq.~(24d) of \cite{canuto98b}, for 
which we take the Kolmogorov constant to be ${\rm Ko} = 1.70$. The turbulent heat diffusivity
$\chi_{\rm t}$ is given by the low viscosity limit of Eq.~(11f) of \cite{canuto98b}. Depending on 
the specific problem molecular dissipation can be included through restoring the largest
second order moment terms containing $\nu$, i.e., $\partial_z (\nu \partial_z K)$, etc.\
They are important when ${\rm Pr}$ is of order unity rather than zero,
\cite{kupka99b}, as is often the case when comparing solutions of convection models 
to numerical simulations for idealized microphysics. Therefore, such a contribution
can also be included in (\ref{eq_epsilon}). For the latter, $c_1=1.44$, $c_2=1.92$, and
$c_3=0.3$ where $\beta < 0$ while $c_3=0$ elsewhere, as suggested in \cite{canuto98b}. 

At this level of complexity the model has been used in \cite{kupka99b,kupka02b,montgomery04b}.
When compared to 3D hydrodynamical simulations it allows accurate predictions for cases with 
large radiative losses, as in sufficiently hot A type and DA type stars (\cite{kupka07f}), but not
for solar-like convection.

\subsection{Two-scale mass flux models} \label{sec-3.4}

For deep convection zones with low radiative heat losses inside the convection zone,
such as in our Sun, a comparison with numerical simulations for idealized microphysics reveals 
\cite{kupka07d,kupka07f} that the models for the TOMs suggested in \cite{canuto93b,canuto01b} 
cannot any more reproduce higher (third) order moments. Those cases of numerical simulations of
compressible, vertically stratified convection for which they had been found to work, such
as discussed in \cite{kupka99b}, had been characterized by small values of skewness
of the vertical velocity and temperature fields.

A large skewness is related to flow topology and results from suitable boundary conditions.
Non-local transport then leads to inhomogeneity of the flow, the formation of an asymmetric
distribution of up- and downdrafts. This has been analyzed in meteorology already about
30 years ago. The physical mechanism behind the asymmetry between top-down and bottom-up 
``diffusion'' due to a turbulent flow was discussed in \cite{wyngaard87b} while the nature of the 
vertical-velocity skewness for the planetary boundary layer was discussed in \cite{moeng90b}.
Narrow convection zones with strong radiative losses reduce the influence of the boundary
layers which leads to a smaller skewness of the vertical velocity and temperature field.
For the atmosphere of the Earth the most detailed in situ measurements of these and related 
quantities has been made during the aircraft campaign ARTIST which took place from 4--9 April
in 1998 (see \cite{hartmann99b}). To explain these data a new model was derived in
\cite{gryanik02b} while a competing model was suggested by \cite{cheng05b}.

In the following an overview on the model by \cite{gryanik02b} is given. From an
analysis of the PBL (planetary boundary layer) aircraft data \cite{gryanik02b} suggested
that coherent structures contributed most to the higher order moments of the vertical velocity
and temperature field and thus also to skewness. They hence considered averages over up-/downflow 
areas and separately averages over hot and cold areas. The result is a {\em two-scale mass flux average}
which replaces the previously suggested {\em mass flux average} (also known as ``elevator model'') to 
compute higher order moments in the {\em ballistic limit of large skewness}. In that case either the up- or the 
downdrafts (or either the hot or the cold regions) are so localized within each horizontal plane that
normal to the latter the fluid moves at high velocity (or large temperature difference) with just limited 
interaction with its environment. This extreme, asymptotic limit for very large skewness gives rise 
to the idea of considering $n$-$\delta$ probability distribution functions for the velocity and temperature
distributions of turbulent flow fields.

The level of averaging required to arrive at this model was visualized through a data record obtained 
from a flight through the convective planetary boundary layer at a certain altitude \cite{gryanik02b}.
Horizontally and in time the unaveraged data vary highly. A top-hat average then segments this function
into constant functions between sign changes for each variable. Here, the sign is obtained from subtracting
the horizontal (and temporal) mean from each quantity measured at a particular vertical depth and instant of time.
Downflow regions have a negative velocity, upflow regions a positive one, and likewise for regions of the
flow where the temperature is lower or higher than the specified average. The {\em two-scale mass flux average}
is then obtained by determining the average over each of the upflow segments, the downflow segments,
the hot flow regions, and the cold flow regions. This is more general than the {\em one-scale mass flux average}
introduced by \cite{arakawa69b,arakawa74b} where an average value is computed for any variable over
all upflow regions joined by a second average which is performed over all downflow regions. An alternative
representation of the two-scale mass flux are 4-$\delta$ probability distribution functions which 
for each horizontal layer describe the distribution as averages over all four combinations of signs
for vertical velocity and temperature. 
For the convective planetary boundary layer, 35\% to 45\% of the total horizontal area are covered
by cold downflows between which an area of 25\% to 30\% is covered by hot, upwards rising plumes
originating from the heating of the flow at the surface. But between 25\% and 40\% are covered
by flow characterized through the other two possible combinations of signs. With the role of signs reversed
(hot upflows dominating in area over cold downflows) this can also be found for solar granulation 
(see \cite{stein98b}), although the detailed distribution varies depending on whether averages are
taken over horizontal planes or planes of constant optical depth. 

Using correlations computed with the 4-$\delta$ probability distribution function to approximate
ensemble averages in the limit of large skewness, the influence of turbulent fluctuations caused
by shear between the up- and downflows is accounted for in a second step. \cite{gryanik02b,gryanik05b} 
suggested these to be obtained from linear interpolation between the quasi-normal, Gaussian limit for the 
case of zero skewness and the ballistic two-scale mass flux limit in the other case. This allows specifying closed 
expressions for fourth order moments of the flow and for cross-correlation third order moments. The skewness 
for both vertical velocity and temperature has to be specified from some other source and the same holds
also for second order moments (estimates for $\overline{w\,\theta}$ from the two-scale mass flux turned 
out to be not useful for improved closure relations, as it is much less correlated to up- and downflow regions
than its higher order counterparts, see \cite{gryanik02b}).
In this form the model was tested with both aircraft data and simulation data for
the convective planetary boundary layer. Root mean square errors between model
and data were found to be drastically smaller than for the quasi-normal approximation. 
The top-hat average two-scale mass flux relations of \cite{gryanik02b} read as follows:
\begin{eqnarray}
  <v_z>_{\rm mf} & = & a\,w_{\rm u}        + (1-a)\,w_{\rm d},                          \label{eq.vzmfa} \\
  <T>_{\rm mf}   & = & b\,{\theta}_{\rm h} + (1-b)\,{\theta}_{\rm c}.                  \label{eq.Tmfa} \\
  <w^2  \theta>_{\rm mf} &=& S_w\,\sigma_w\,<w \theta>_{\rm mf},               \label{eq.w2tmfa} \\
<w  \theta^2>_{\rm mf} &=& S_{\theta}\,\sigma_{\theta}\,<w \theta>_{\rm mf}, \label{eq.wt2mfa} \\ 
<w^4        >_{\rm mf} &=& \left(1+S_w^2 \right) \sigma_w^4,                         \label{eq.w4mfa}  \\
<   \theta^4>_{\rm mf} &=& \left(1+S_{\theta}^{2} \right) \sigma_{\theta}^4,      \label{eq.t4tfa}  \\
<w^3\theta  >_{\rm mf} &=& \left(1+S_w^2 \right) \sigma_w^2\,<w\theta>_{\rm mf},      \label{eq.w3tmfa} \\
<w  \theta^3>_{\rm mf} &=& \left(1+S_{\theta}^2 \right) \sigma_{\theta}^2\,<w\theta>_{\rm mf}. \label{eq.wt3mfa} 
\end{eqnarray}
The quantities $w_{\rm u}$ and $w_{\rm d}$ are the horizontal averages over {\em all} 
up- and downflows relative to the mean upwards velocity, $<v_z>_{\rm mf}$. In the
same sense, ${\theta}_{\rm h}$ and ${\theta}_{\rm c}$ are the horizontal averages over
all areas with a temperature higher and lower than the mean temperature $<T>_{\rm mf}$,
respectively. For the two-scale mass flux averages the external parameters 
$a$, $b$, $w_{\rm u}-w_{\rm d}$, and ${\theta}_{\rm h}-{\theta}_{\rm c}$ can be replaced
self-consistently by the quantities $\sigma_w$, $\sigma_{\theta}$, $<w\theta>_{\rm mf}$,
$S_w$, and $S_{\theta}$. This has already been used here for Eq.~(\ref{eq.w2tmfa})--(\ref{eq.wt3mfa}).

From this starting point \cite{gryanik02b,gryanik05b} have suggested that the ensemble averages
\begin{eqnarray}
\overline{w^2 \theta} & = & S_w\,\sigma_w\,\overline{w \theta} = 
    \left(\overline{w^3}/\overline{w^2}\right)\,\overline{w \theta},                                                   \label{eq.w2t} \\
\overline{w \theta^2} & = & S_{\theta}\,\sigma_{\theta}\,\overline{w \theta} = 
    \left(\overline{\theta^3}/\overline{\theta^2}\right)\, \overline{w \theta},                                       \label{eq.wt2} \\
\overline{w^4        } &=& 3 \left(1+\frac{1}{3} S_w^2 \right)          \overline{w^2}^2,                       \label{eq.w4gh}  \\
\overline{   \theta^4} &=& 3 \left(1+\frac{1}{3} S_{\theta}^{2} \right) \overline{\theta^2}^2,               \label{eq.t4gh}  \\
\overline{w^3\theta  } &=& 3 \left(1+\frac{1}{3} S_w^2 \right) \overline{w^2}\,\overline{w\theta},      \label{eq.w3tgh} \\
\overline{w  \theta^3} &=& 3 \left(1+\frac{1}{3} S_{\theta}^2 \right) \overline{\theta^2}\,\overline{w\theta}, \label{eq.wt3gh} \\
\overline{w^2\theta^2} &=&   \overline{w^2}\,\overline{\theta^2} + 2 \overline{w \theta}^{2}
		               + S_w\,S_{\theta}\,\overline{w \theta}\,\sigma_w\,\sigma_{\theta},                 \label{eq.w2t2gh}
\end{eqnarray}
hold in the limit of both large and small skewness. Here, the cross-correlation for Eq.~(\ref{eq.w2t2gh})
has been added which has been derived in \cite{gryanik05b} to complete the closure relations of 
\cite{gryanik02b}. The relations for Eq.~(\ref{eq.w2t})--(\ref{eq.wt2}) had already been suggested in earlier 
work (see \cite{zilitinkevich99b,mironov99b}). Note that for zero skewness this new model 
coincides with the Millionshchikov 1941 (quasi-normal) approximation \cite{millionshchikov41b}. 
Alternatively, it was suggested in \cite{gryanik02b} to consider optimized parameters $a_i$ and $d_i$ 
instead of 3 and 1/3. The improvement gained from such a procedure is much smaller though than that 
one already gained by Eq.~(\ref{eq.w2t})--(\ref{eq.w2t2gh}) relative to the QN, \cite{millionshchikov41b}.

The realizability of the closure approximations Eq.~(\ref{eq.w2t})-(\ref{eq.w2t2gh}) was demonstrated in
\cite{gryanik05b} for the planetary boundary layer using aircraft data and numerical simulations as well as 
in \cite{kupka07b} for granulation in the Sun and in a K-dwarf using hydrodynamical  simulations. The 
quasi-normal approximation yields non-realizable results for the temperature field of the atmosphere 
and for both the velocity field and the temperature field for both types of stars.

Studies which show the substantial improvement of either some or all of Eq.~(\ref{eq.w2t})-(\ref{eq.w2t2gh}) 
when compared to previous models include a resolved numerical simulation of free oceanic convection, 
\cite{losch04b}, the numerical simulation of deep, compressible convection with idealized microphysics, 
\cite{kupka07c,kupka07d,kupka07e,kupka07f}, the numerical simulation of solar granulation, 
\cite{kupka07b,kupka09c,kupka17b}, and granulation in a K-dwarf, \cite{kupka07b}, as well
as in a DA white dwarf, \cite{kupka17b}, and again simulations of compressible convection with 
idealized microphysics, \cite{cai18b}. 

Given these credits, what are the known limitations of this model apart from still existing (though smaller) 
discrepancies between direct computations of the ensemble averages and their approximation through 
Eq.~(\ref{eq.w2t})-(\ref{eq.w2t2gh}) ? Clearly, the up- and downflows trigger shear induced turbulent 
fluctuations at their boundaries (see \cite{stein00b} and also \cite{muthsam10b}). In the model of 
\cite{gryanik02b,gryanik05b} a quasi-normal distribution is assumed to hold for the case of zero skewness of 
a field of fluctuations of velocity or temperature. But solar granulation provides a counter example: at the bottom
of the solar superadiabatic peak the skewness $S_w = \overline{w^3} / (\overline{w^2})^{1.5}$ drops from $\approx -0.2$ 
to $-1$ whereas the kurtosis $K_w = \overline{w^4} / (\overline{w^2})^2 \approx 2.5$ instead of $3$ in that region. 
Likewise, $S_{\theta} = \overline{\theta^3} / (\overline{\theta^2})^{1.5} \approx 0$ there, whereas the kurtosis 
$K_{\theta} = \overline{\theta^4} / (\overline{\theta^2})^2 \approx 1.3$ instead of~$3$. Values of $K_{\theta} \approx 2$
at the bottom of the surface superadiabatic peak are also observed for numerical simulations of 
DA white dwarfs, \cite{kupka18b}, or K dwarfs, \cite{kupka07b}, and independently of vertical boundary 
conditions or the particular simulation code used. At the same time the large values of the kurtosis deep
inside thick convection zones are still somewhat underestimated by the model of \cite{gryanik02b,gryanik05b}
(see again \cite{kupka07b}). Both the modelling of statistical properties of the large scale coherent 
structures and the local turbulence created by the flow still require further improvement, although
the discrepancies are already much lower, typically by factors of 2 to 3, than those obtained with the 
quasi-normal approximation. Plume models for the downdrafts may provide some of these improvements
(see \cite{belkacem06b}). Recalling the Reynolds stress models it might seem attractive to combine 
the closures suggested in \cite{gryanik02b,gryanik05b} to a more general model (see  \cite{gryanik02b}).
However, the naive combination of ``favourite closures'' including those by \cite{gryanik02b,gryanik05b}
easily triggers instabilities as has been shown in \cite{kupka07f} and confirmed by \cite{cai18b}.

\subsection{Comparisons} \label{sec-3.5}

For systems with efficient convection the comparison with numerical simulations 
implies that the non-local convection models based on the downgradient approximation
for third order moments or the quasi-normal approximation of fourth order moments
(such as those used in \cite{kupka99b,kupka02b,montgomery04b}, but also \cite{xiong86b,xiong97b}
or \cite{kuhfuss86b}) cannot reproduce third order moments and thus non-local transport effects 
(cf.\ \cite{kupka07c,kupka07d,kupka07e,kupka07f}) despite such models claim to account for these
effects. The cases of interest characterized by efficient convection all feature large absolute 
values of skewness which have to be a consequence of boundary conditions and non-locality
(see \cite{wyngaard87b} and \cite{moeng90b}, e.g.). This leads to a strong spatial inhomogeneity 
of the flow with strong asymmetric up- and downflow areas, as has also been found for geophysical
cases such as convection in the atmosphere of the Earth or in the ocean.

But what about cases in which such models have been shown to work, as in 
\cite{kupka99b,kupka02b,montgomery04b} or in \cite{canuto01b}? Can they
be distinguished from the other cases? One important difference is that these
``good cases'' all have small absolute values of skewness. Hence, it is less important
that coherent structures and flow topology have not been taken into account in much
detail in these convection models. In turn, convection models that explicitly account 
for flow structure are clearly favourable over those which do not, as these features 
are an essential part of the physics of convection.

\newpage

\section{Challenges and pitfalls in numerical modelling}  \label{sec-4}

Hydrodynamical simulations are now a key tool of stellar astrophysics. 
A summary on codes and simulation strategies for modelling convection has been
given elsewhere, \cite{kupka17b}. The topics highlighted in the following are usually 
paid less attention for in discussions of numerical modelling of solar and stellar convection. 
Their closer consideration should help avoiding some of the main pitfalls and allow a better 
understanding of the challenges of this approach.

\subsection{General remarks} \label{sec-4.1}

Due to finite computational resources any hydrodynamical simulation of stellar convection 
for the case of realistic microphysics and realistic radiative fluxes has to consider a volume 
averaged approximation of $\rho(t,{\bf x}), \rho {\bf u}(t,{\bf x}), \rho e(t,{\bf x})$. In the language
of engineering sciences these are hence {\em large eddy simulation (LES)}. As has already been
explained in Sect.~\ref{sec-3.2} in that case the numerical simulation directly resolves the scales 
carrying most of the kinetic energy (e.g., scales on which radiative cooling takes place at the stellar 
surface) on the computational grid. Smaller scales are taken care of by some {\em subgrid scale model}. 
Often this may just be some numerical or artificial viscosity. Scales larger than the computational
grid have to be handled by boundary conditions. In practice the simulation is started from 
``typical'' initial conditions and {\em relaxed} towards a {\em statistical equilibrium} state (see Sect.~\ref{sec-3.2}).

A statistical interpretation of the results of a numerical simulation of convection is thus
part of the computational procedure. The time integration of the simulation is used to generate 
both ensembles and the {\em typical cases} (shown as snapshots such as Fig.~\ref{fk.fig-1}). 
The computation of averaged values over sufficiently long time intervals allows a comparison 
with observations, for instance of spectral absorption lines, and the computation of direct ensemble 
averages. Other tools used for the interpretation of hydrodynamical simulations include also graphical 
visualization or interpretation of the simulation data in Fourier space. 

How well does this concept deal with turbulent flows in practice? To answer this question 
some terminology has to be introduced first. For the {\em phenomenology of turbulent flows}
there is no commonly accepted definition. In a strict sense (\cite{tsinober09b}) it refers to anything 
but direct experimental results, direct numerical simulations, and a few results obtainable from first principles. 
From this viewpoint any, necessarily large eddy type, simulation of stellar convection with realistic
microphysics would be considered phenomenology. However, in astrophysics the usage of the word 
phenomenology is restricted to models that introduce a concept such as rising and falling 
bubbles which cannot be derived from first principles nor confirmed by experiment or numerical simulation,
but is used in deriving mathematical expressions of the model. Stellar mixing length theory
is a classical example for that.

Turbulent flows are distinguished by being capable to give rise to large scale, coherent structures
in spite of the long-term unpredictability of the flow itself. The claim that statistics and structure contrapose 
each other is among the common misconceptions about turbulent flows, as explained in \cite{tsinober09b}.
Indeed, the conduction and evaluation of numerical simulations of (turbulent) flows are based on assuming 
statistical stationarity and quasi-ergodicity from the beginning. Thus, any numerical simulation of this
kind is based on a statistical approach to the physical problem.

Of key importance in this context is the {\em quasi-ergodic hypothesis}. It is frequently worded as follows:
{\em ``The time average of a single realization, which is given by one initial condition, is equal to an average 
over many different realizations (obtained through different initial conditions) at any time $t$ in the limit of 
averaging over a large time interval and a large ensemble.''} (cf.\ \cite{tsinober09b}). For an introduction into
statistical descriptions and ensemble averages of turbulent flows see Chap.~3 of \cite{pope00b}.

This hypothesis is fundamental to any numerical simulation of stellar convection and one consequence
thereof is that the detailed initial conditions on the velocity fields are ``forgotten'' by the flow after a relaxation
time $t_{\rm rel}$. This cannot be proven to hold for all flows: there are known counterexamples, but for some flows 
such as statistically stationary (time independent), homogeneous (location independent) turbulent flows 
it can be corroborated even directly from numerical simulations (Chap.~3.7 of \cite{tsinober09b}).

\subsection{Uniqueness of numerical solutions} \label{sec-4.2}

The Navier-Stokes equations depend on non-linear fluxes which are purely
algebraic combinations of the basic, dependent variables and of pressure.
In Eq.~(\ref{sec-2-eq_cont})--(\ref{sec-2-eq_energy}) they appear on the
left-hand side as objects on which the divergence operator acts. Together
with the pressure gradient term in Eq.~(\ref{sec-2-eq_NSE}) contained in
the first term on the right-hand side (separated in Eq.~(\ref{sec-2-eq_NSE-std})
from contributions containing viscosity) they provide the {\em hyperbolic part} of 
the hydrodynamical equations. They are also contained in the Euler equations of 
hydrodynamics which are distinguished from the Navier-Stokes equations by assuming 
zero viscosity ($\eta = 0$, $\zeta = 0$), whence from Eq.~(\ref{sec-2-eq_visc-tensor})
we obtain $\mbox{\boldmath$\pi$}=0$ for this case.

One may wonder why one should not ignore viscosity altogether from the very beginning.
The reason for this idea is that in stars the kinematic viscosity $\nu$ is ``small''. This 
is suggestive for neglecting terms containing it and thus to just solve Euler's equations 
instead of the full Navier-Stokes equations (NSE). 

Indeed, if one numerically approximates the corresponding terms in the NSE, 
the resulting contributions are extremely small. If one decides to just not care about viscosity 
and solve the Euler equations {\em without any viscosity}, one has to face an unpleasant problem:
the resulting solutions can {\em violate the fundamental laws of thermodynamics} and 
there can be even an {\em infinite number of solutions to a given initial condition}.

A unique, physically consistent solution is needed instead. At the root of this problem
is the fact that {\em hyperbolic conservation laws can have discontinuous solutions developing 
in finite time for smooth initial conditions}. {\em Strong solutions} (as obtained for the 
partial differential equation form of a conservation law) are unavailable in that case
while {\em weak solutions} (of the integral form of a conservation law) can still be found.
Already for the special case of the Riemann problem one can find out about and understand 
the consequences of these properties. In the Riemann problem initial conditions with 
a jump between two constant states and different initial velocities in those regions are
considered. If those initial conditions have characteristics (special invariants of the solution) 
going into discontinuity, a unique solution exists for that problem. If the characteristics are instead
going out of it, even an infinite number of solutions is possible. Many of these solutions
will be unphysical. But {\em a unique physical solution exists} for fairly general cases (see \cite{quarteroni94b}),
if one requires the solution to be that one of a more general, second order partial differential
equation for which viscosity has not been neglected yet and {\em takes the limit $\nu \rightarrow 0$}.

This solution is known as the ``{\em entropy solution}'': it is consistent with thermodynamics and it exists 
for the Euler equations if they are interpreted as limit $\nu \rightarrow 0$ of the Navier-Stokes equations.
For the Euler equations they can be obtained directly from imposing the {\em Rankine-Hugoniot (jump) 
conditions} at discontinuities of the solution. This ensures the uniqueness of the solution and as a result 
the conservation laws also hold across solution discontinuities. Finally, consistency of this solution 
with the laws of thermodynamics is ensured. Consequently, {\em all} numerical simulation codes 
(and in particular any LES of stellar convection) must use some prescription of viscosity. 
Examples include {\em numerical viscosity}, {\em artificial viscosity}, and others. 
Its contribution is often difficult to extract from simulations which can become inconvenient
if one has to compute the dissipation rate of (turbulent) kinetic energy or related quantities.
It is important to note that this form of viscosity is required in addition to others which 
are introduced to account for unresolved scales as in any LES of a turbulent flow 
(such as {\em hyperviscosity} or {\em subgrid-scale viscosity}, ...). The inevitability of having
to introduce viscosity makes any claims of {\em parameter freeness} of numerical simulations
of convection for systems with small viscosity pointless. This has been discussed in detail
in Chap.~6 of \cite{kupka17b} and is briefly addressed below in Sect.~\ref{sec-4.5}.

\subsection{Initial conditions and relaxation} \label{sec-4.3}

Numerical simulations of convection have to start from some initial state of the system.
A ``statistically characteristic'' initial condition for the given problem, say a simulation
of solar granulation, cannot be constructed analytically from first principles. Rather,
some more idealized state will have to be considered instead, such as the horizontal
average given by a one-dimensional stellar structure model. In the first part of the
numerical simulation the system is evolved until it {\em forgets} this peculiar initial state.
Data for detailed analysis are collected only for {\em relaxed states} of the simulation. 

For the case of numerical simulations of the solar surface which are based on small boxes 
with Cartesian geometry ({\em box-in-a-star simulations}) the relaxation process was studied
in more detail by \cite{grimm-strele15b}. Three solar structure models based on
two different stellar evolution codes with two different convection models being used
provided initial conditions for three numerical simulations distinguished
from each other only by their initial state. Except for a very small difference caused
by slightly different entropies of each initial condition the simulation averages
after one hour of solar time turned out to converge to the same solution. This was
tested by comparing entropy and the superadiabatic gradient as a function of depth
and of pressure (see Fig.~9 and 10 in \cite{grimm-strele15b}). As the initial states
differ by more than 1\% only for the superadiabatic peak and for the photosphere, 
this is not surprising: the thermal time scale for those layers is less than one solar hour. 
Once these layers are statistically relaxed and since the region underneath them 
is quasi-adiabatic with just a small difference in entropy between each of the 
three solar structure models, the simulations are thermally relaxed and thus 
have the same mean thermal structure. This is discussed in further detail
in \cite{kupka17b}. In \cite{kupka18b} relaxation is studied for
a DA white dwarf with a shallow surface convection zone. 

The good agreement for solar granulation simulations made with different 
numerical simulation codes (cf.\ \cite{kupka09c,beeck12b}) is thus not
a complete surprise: as long as resolution, microphysics, chemical composition,
and the entropy in the quasi-adiabatic layers are comparable, relaxation
proceeds to a similar physical state after just one solar hour. This is the
thermal relaxation time for the superadiabatic zone and also the convective
turnover time scale which guides relaxation of kinetic energy, as it is tied
to the entire energy being transported by the flow, i.e., the convective motions.

The question of {\em how long do we have to relax a simulation of convection} before proceeding with 
its statistical analysis can thus be answered as follows \cite{kupka17b}: relaxation has to ensure that
the mean stratification is in thermal equilibrium. Thus, $t_{\rm rel} \sim t_{\rm therm}$ at least.
Since this can become computationally very expensive, the main trick for {\em fast relaxation} when
doing two- and three dimensional (2D and 3D) numerical simulations of stellar convection is to guess 
a thermally relaxed stratification from a proper one-dimensional model. For simulations of the
top layers of a deep, quasi-adiabatic convection zone such as our Sun this is quite easy. There,
$t_{\rm rel} \sim t_{\rm KH}(r)$, the local, depth dependent Kelvin-Helmholtz time scale
(see both \cite{grimm-strele15b} and \cite{kupka17b}). This is the time it takes to exchange the
thermal energy contained in the layers above $r$ with the environment for a fixed luminosity or input flux. 
Only if the dominating relaxation process is radiative diffusion, then $t_{\rm rel} \sim t_{\rm diff}(r)$ instead. 
Here, the timescales are based on integrals ranging from the top of the simulation to the layer $r$ below 
which the mean stratification remains roughly constant with time due to the layers underneath being relaxed 
from the beginning. Since no good initial condition can be guessed for the velocity field, the relaxation of 
kinetic energy requires $t_{\rm rel} \gtrsim t_{\rm conv}(r_{\rm bottom})$. Generally, 
\begin{equation}
t_{\rm rel} \sim \max(t_{\rm conv}(r_{\rm bottom}), t_{\rm KH}(r_{\rm rel}))
\end{equation}
with an initial condition chosen which hopefully permits $r_{\rm rel} > r_{\rm bottom}$. 

In \cite{grimm-strele15b} it was also shown that the {\em relaxation of kinetic energy} may
take much longer for a 2D simulation of stratified (compressible) convection than for a 3D one.
The 2D case may show a plateau in kinetic energy for some time (see their Fig.~12), but then 
this quantity can start to increase or decrease again. This is likely connected to the fact
that in 2D large scale (vertically oriented) vortices form with a long life time and their merging
and disappearing and reappearing may cause major changes in the amount of kinetic energy
in the flow. As discussed in \cite{tsinober09b}, 2D flows are very often non-ergodic even when
their 3D counterparts are quasi-ergodic. Hence, their relaxation and statistical evaluation may 
set requirements quite different from those of 3D simulations for the same scenario.

Successful relaxation then sets the stage for further analysis of the results of numerical simulations 
of convection. How long do we have to average and evaluate a simulation?
For time independent, quasi-stationary quantities, this is determined by the number of realizations 
required to obtain statistical stationarity and $t_{\rm stat}$, the time scale over which we want to 
do averages to compute statistical properties, may hence vary from a few snapshots to many hundreds 
of sound-crossing times.

In fact $t_{\rm stat}$ can also depend on the width of the simulation box or, equivalently, on the number of 
granules contained in it. If the quantity of interest is truly time dependent, $t_{\rm stat}$ depends on the 
time scales of the underlying physical process such as stellar oscillations, mode growth rates, or mode damping 
time scales. Provided sound waves and radiative transfer are also properly taken care of, 
$N_{\rm tot} = (t_{\rm rel} + t_{\rm stat}) / (\delta t_{\rm adv}) \sim 10^6 \dots 10^8$ time steps can be afforded 
for a 3D simulation of stellar convection with about $400^3$ grid points. As an order of magnitude estimate
this separates computable from unaffordable problems on top department clusters (on top supercomputer clusters 
calculations which are some two orders of magnitudes more expensive are typically still affordable, see Sect.~\ref{sec-4.6}). 
If this is to be circumvented, radiative transfer has to be accelerated, local resolution has to be reduced, or other 
measures have to be taken to proceed with a computing project in acceptable time.

\subsection{Boundary conditions} \label{sec-4.4}

Vertical boundary conditions in simulations of stellar convection are parametrized. The reason for
this is that the {\em inflow of mass and energy} at the vertical boundaries has to be modelled. Even without 
explicit numerical constants involved, this inevitably requires some form of parametrization. The most simple 
assumption is that of a closed vertical boundary with zero vertical velocity. Since that keeps the fluid
completely inside the computational box, any inflow and outflow of energy has to be due to
radiation (or artificial conduction). If inflow and outflow of fluid are permitted, the parametrization
becomes more delicate. 

Due to the extreme stratification of the fluid and a comparatively moderate flux of energy through 
stellar convection zones, a feature illustrated by the ratio of kinetic and potential energy in simulations
of solar granulation, \cite{grimm-strele15b}, the mean thermal structure of stellar convection zones is
quite robust. The mean temperature structure in the superadiabatic layer is even independent of whether
closed (impenetrable and reflecting) boundaries or open boundaries have been used for the vertical velocity 
field in published simulations (see \cite{robinson03b,kupka09c}).

However, this is not the case for many higher order correlations. Especially, quantities depending on 
horizontal velocities are sensitive (cf.\ \cite{robinson03b}), but also higher order correlations which 
depend on the temperature field and vertical velocity field (see \cite{grimm-strele15b}).
A safety margin of between $1$ and $2$ local pressure scale heights as measured at each 
boundary is needed even in case of the best vertical boundary condition models currently 
available (cf.\ \cite{kupka07b,grimm-strele15b}). Evidence for this comes from dedicated 
3D numerical simulations of solar granulation made with the ANTARES code which differ
only in the boundary conditions chosen (see \cite{grimm-strele15b}).
There, the mean temperature was found to be invariant to variations of the vertical 
boundary conditions. However, the flux of kinetic energy turned out to be sensitive 
to those changes in the entire lower half of the simulation domain (of about 5~Mm), 
despite the vertical boundary conditions considered all permitted free outflow and 
an inflow, as specified by the particular boundary condition model. Closer inspection 
demonstrated that also the probability distribution of vertical velocities is equally 
sensitive to details of the boundary conditions. In the superadiabatic layer the velocity 
distribution was found insensitive to the boundary conditions applied several pressure 
scale heights away from it. However, half-way to the bottom of the simulation box, 
some 2~Mm below the optical surface, systematic difference are apparent which close 
to the bottom become striking \cite{grimm-strele15b}. The recommendation drawn from 
this was to opt for the boundary condition causing the smallest change in the shape of the velocity
distribution as function of depth. This choice also minimizes spurious features in the fluxes 
of kinetic energy and enthalpy and other quantities such as skewness and kurtosis.

\subsection{Criteria for modelling replacing the quest for parameter freeness} \label{sec-4.5}

The task of choosing the ``best'' vertical boundary condition motivates the question which 
criteria have to be applied when choosing parameters in any approach to model convection. 
The parameters to be worried about do not include those directly linked to the physical system 
studied (mass of a star, radius, etc.), but those introduced by physical and numerical 
approximations. Against widespread opinion numerical simulations have this problem, too.

The quest for a parameter free modelling of convection has a long tradition in astrophysics.
In the end the only way to completely avoid ``free parameters'' is to consider a flow field so 
idealized as to become unrealistic. Examples include assuming the flow field is
\begin{enumerate}
\item irrotational and allows the Boussinesq approximation. Then, a closed model can be 
        derived, \cite{pasetto14b}, without the need to specify closure constants.
        This approach completely suppresses turbulence of the flow and can hold 
        only for small contrasts in density.
\item A two-point Dirac distribution function describes the vertical velocity field. This is the idea
        behind the mass flux average, \cite{arakawa69b,arakawa74b}, which averages quantities 
        separately over up- and downstreams. This yields exact closures for higher order correlations
        but ignores small scale turbulence, pressure fluctuations, among others.
\end{enumerate}
Such assumptions can at best give hints to new closure approximations, but for applications
to real world flows those again have to be parametrized.

Less commonly known are the unavoidable parameters introduced in 2D and 3D numerical
simulations of convection, \cite{kupka17b}. These include
\begin{enumerate}
\item mathematical ones (time integration, spatial discretization);
\item the viscosity model (subgrid-scale models, hyperviscosity, artificial diffusion, numerical viscosity);
\item and the boundary conditions.
\end{enumerate}
The key point to be remembered is that {\em the mere existence of model parameters is not crucial}.
What actually matters is the {\em nature of the entire model (completeness of physics included) 
and how the parameters are adjusted}.

How is this problem solved in the case of numerical simulations of convection?
\begin{enumerate}
\item For the mathematical ones, optimization procedures are performed by mathematicians
        based on generic criteria (stability, etc.). {\em Any retuning later on should be avoided}.
\item The viscosity model is calibrated with idealized test problems (shock tubes, etc.), 
        {\em but not with astrophysical ones and should not be retuned later on}.
\item The boundary conditions are calibrated with single or a few astrophysical test cases, 
        {\em but once fixed should not be altered any more}.
\end{enumerate}

This motivates an alternative approach to the problem of {\em free model parameters}, \cite{kupka17b}.
The idea is to define properties that both numerical simulation based models and any local or non-local 
model of convection should have and use this as a set of criteria replacing the (unrealistic and not very helpful) 
quest for parameter freeness. Those criteria were suggested and discussed in detail in \cite{kupka17b}.
The model or approximations it contains should
\begin{enumerate}
\item have the {\em correct physical dimension};
\item ensure {\em invariance of tensor properties} and proper behaviour with respect to standard
        transformations (coordinate systems);
\item respect {\em sign- and other symmetries} of basic variables;
\item allow physical and mathematical {\em realizability};
\item ensure {\em robustness}, i.e., predictions should be robust with respect to reasonable changes 
        of the parameters or replacement of the model component containing them by one 
        which is robust in this sense, too;
\item allow for {\em universality}, i.e., it should be unnecessary to recalibrate the internal parameters 
        for different types of astrophysical objects (e.g., Sun, DA type white dwarfs, etc.);
\item permit {\em computability}, i.e., the formalism must be affordable on present computing means;
\item allow {\em physical verifiability}. This requires the model to be falsifiable with observational
        data or direct numerical simulation or an approach having passed such tests;
\item ensure {\em independence of its internal parameters from the object being modelled}.
\end{enumerate}

\subsection{Computability} \label{sec-4.6}

Due to the enormous increase in computer power during the 
last five decades hydrodynamical simulations of astrophysical flows 
are now a standard tool of research in all fields of astrophysics. 
The success in solving some problems including those involving 
stellar convection has spread hope to apply this tool ever more widely.
It has hence become a common expectation in astrophysics in general and 
in stellar astrophysics in particular that {\em ``we just need to wait until 
computers are powerful enough to perform a 3D simulation''} of a given problem.

Thus, also studies of stellar convection, stellar pulsation, and stellar evolution 
are confronted with this demand. Sometimes this takes place during discussions 
at scientific conferences where it certainly is raised as a legitimate and interesting 
question. However, the same question or demand can change into a serious issue,
if it is voiced by peer-reviewers of research publications or even grant applications.

It is thus important to become aware that {\em some problems can or will be solved 
this way. However, others cannot, at least not with semiconductor based technology.}
In \cite{kupka17b} a detailed analysis of several examples from stellar convection 
studies has been given in which some problems had been concluded to be 
{\em ``computable''} in this sense while others were not. 

It is important to be aware that this computability analysis is not about ``exact'' predictions, 
since details vary depending on software and coding efficiency, numerical methods, 
hardware platforms, and the like. However, for the main conclusions this has limited 
impact: {\em presently doable problems and their requirements are known. They can be
distinguished from problems for which this is not the case yet and from problems for 
which this will not be the case for unforeseeably long times to come.}

The detailed estimates have already been given in \cite{kupka17b}. For convenience
and to avoid misunderstandings the underlying assumptions are summarized 
here along with some additional remarks.
First and foremost, a hydrodynamical simulation of stellar convection is time step limited 
by $\tau \sim \Delta t_{\rm adv}$. This is the time scale on which fluid is advected between
neighbouring grid cells. That restriction cannot be circumvented by implicit methods which 
is possible for constraints such as those caused by sound waves or rapid radiative cooling: 
those may operate rapidly but they do not have to be followed in detail on time scales on which
they do not cause a large relative change of the solution (see \cite{kupka17b}). 
Since a spatial resolution $h \lesssim 12\,{\rm km}$ is needed to obtain a properly resolved 
solar superadiabatic layer, there $\tau \sim \Delta t_{\rm adv} \sim 1\,{\rm sec}$.

Typical moderately high resolution 3D simulations done today have $\sim 400$ grid points 
per direction. To relax and average the simulation at such a time step requires 
$N_t = (t_{\rm rel} + t_{\rm stat})/(\Delta t_{\rm adv}) \sim 10^6$ time steps with 
$N \sim 10^8$ grid points. This is readily solvable on present supercomputers and has been 
assigned a reference computational complexity of $C:=1$. The estimate is also based on
a {\em thermally relaxed initial solution for the adiabatic interior} (\cite{kupka17b}).

A similar estimate can be obtained for simulations excluding the stellar surface, accounting for cases of 
implicit integration of radiative diffusion or sound waves (or filtering), or solving the radiative transfer equation 
using binned opacities. The variation of $C$ caused by these changes is less than about a factor of 10.
In 2D simulations one can trade grid points for time steps to arrive at similar values of $C$. Grid refinement methods
can gain another factor of $\sim 10$. Optimized parallelization (including GPUs) may gain factors anywhere
between 10 and 100 (this is naturally the least certain part of the estimate). At the bottom line though this allows 
sufficiently robust estimates to distinguish doable from undoable problems. 

Problems doable or becoming doable today have $C \lesssim 10^3$ and include well-resolved simulations of solar granules in 
a 6~Mm wide (and about 4~Mm deep) box, a fully turbulent simulation of the same problem ($h \lesssim 3\,{\rm km}$), low
resolution simulations of very large and deep boxes (200~Mm wide and deep), subsurface simulations of global solar 
convection, and 2D simulations of short period Cepheids that include their first interior node (42\% of the stellar
radius from the top) for both narrow wedges and full azimuth covering simulations. 

With the upcoming generation of exascale computing systems problems with a complexity of $C$
in the range of $10^4$ to $10^6$ and perhaps even $10^7$ might become doable. This includes 3D
simulations of the entire solar surface or 3D simulations of Cepheids (both have $C \lesssim 10^5$). 
The limiting factor for the next step, a low resolution simulation of the entire solar convection zone, is 
then to find a compromise between how low the resolution of the surface layers can still become to yield 
acceptable results and the question to what extent one can accept simulations based on kinetic relaxation
only. The example given in \cite{kupka17b} refers to a problem with a vertical resolution of 12 to 28~km 
for the surface layers and a kinetic relaxation over one year resulting in $C > 10^7$ which seems
unlikely to be achievable shortly. It is of course possible to downscale this computation with respect to
resolution and to push the problem into the realm of simulations which will become affordable with new codes and 
new hardware expected to be developed and released during the upcoming decade. However, a 3D simulation
of the entire solar convection zone that would be usable to study damping of p-modes would be much
more demanding, since downscaling of relaxation is not an option there while simulated periods have
to cover one or, rather, several decades. A technology suitable to cope with $C > 10^{10}$ appears
unforeseeable as of now. One might of course speculate about the capability of quantum computers
in the future, but this cannot be linked to a technological development path with predictable production
cycles. This restriction also includes low resolution 3D simulations of stellar evolution for any foreseeable 
time ($C > 10^{17}$). The only option in that field is to rather use numerical simulations to calibrate
advanced 1D models.

\subsection{A comparison} \label{sec-4.7}

The complex challenges of calibrating advanced 1D models of convection with numerical
simulations can be elaborated when considering the case of Cepheids \cite{mundprecht15r}.
The standard 1D non-local models of convection used to quantify the convective flux
in radial simulations of Cepheid pulsation require the specification of closure parameters
which change as a function of time, pulsation phase (since the mean stratification varies
substantially contrary to the case of solar-like oscillations), and location inside the 
convection zone or in the overshooting region. In the case of a short-period Cepheid
($P_{\ast} \sim 4\,{\rm d}, M_{\ast} \sim 5\,M_{\odot}, R_{\ast} \sim 38.5\,R_{\odot}, L \sim 913\,L_{\odot}$)
one of the parameters of either convection model studied (\cite{stellingwerf82b,kuhfuss86b})
was found to have to be varied by at least a factor of 2 as a function of each of the
dependent quantities considered (i.e., pulsation phase, location inside the convection zone, etc.).
Evidently, reliably calibrating such models is a highly non-trivial issue. With respect to
the criteria listed in Sect.~\ref{sec-4.5} one might thus compare the different approaches
to modelling convection as follows. 

The requirements of dimension, invariance, and symmetries are generally fulfilled by the 
models in active use and definitely for all well-tested numerical simulations (some closure 
approximations may violate sign symmetries though). Realizability is mostly an issue
for some of the more advanced 1D models of convection. However, the other five criteria
clearly divide the different modelling concepts. Classical mixing length theory benefits
from being the cheapest computationally and can be (and indeed has been) falsified in
astrophysical applications. It performs poorly though with respect to robustness of its
results, universality, and the demand of object independence. This is actually the real
problem behind adjusting the mixing length in astrophysical applications, especially,
if this is done based on astronomical observations. At increased computational
costs the non-local models of convection and in particular the Reynolds stress models
aim at improving on those short-comings. The extent to which this is possible remains
a part of active research and depends largely on the class of objects and the physical
process chosen to be modelled with the desired level of accuracy. Numerical simulations
on the other hand perform well for each of the criteria except for computability, as can
also be concluded from the discussion in Sect.~\ref{sec-4.6}.

For that reason in stellar astrophysics and in related fields in planetary science one has
to cope with a larger variety of concepts to model convection for the foreseeable future. 

\newpage

\section{Applications: overshooting}  \label{sec-5}

One important problem in the modelling of thermal convection in stars
is to quantify the amount of mixing next to an existing convection zone. 
Some results on this process can be obtained in a fairly general 
form through considering balanced budgets of ensemble averages, 
even on the level of Reynolds stress models. After introducing
such more general principles the problem of overshooting in white
dwarfs and the role of the P\'eclet number in characterizing overshooting
are highlighted.

\subsection{Overshooting} \label{sec-5.1}

In the following the {\it convective planetary boundary layer} (PBL) of the atmosphere 
of the Earth is considered as a model system. For this case one can explicitly show 
that using an analysis just based on Reynolds stress modelling it is possible to 
demonstrate that convective overshooting, the penetration of fluid from a zone 
convectively unstable according to the Schwarzschild criterion into neighbouring 
layers which are stably stratified from that point of view, is an inherently non-local
phenomenon. Advective fluxes play a key role in establishing energy transport
in an {\em overshooting zone} and determine its basic physical properties.

We start from the Reynolds stress equations for the second order moments introduced in Sect.~\ref{sec-3.3}
(cf.\ \cite{canuto92b}, \cite{canuto98b},  \cite{kupka99b}): {\em turbulent kinetic
energy} $K$,  {\em root mean square fluctuations of temperature} $\overline{\theta^2}$, and their 
cross relation $\overline{w\theta}$ (essentially, the  {\em convective flux}). They read
\begin{eqnarray}
 \lefteqn{\partial_t K + \partial_z \left(\frac{1}{2}\overline{q^2 w} + \overline{p w}\right)  =  
         g \alpha_{\rm v} \overline{w\theta} - \varepsilon} \nonumber\\
        & & {} + \partial_z \left(\nu \partial_z K\right) + \frac{1}{2}C_{ii},   \label{Kfull}
\end{eqnarray}
\begin{eqnarray}
 \lefteqn{\partial_t \left(\frac{1}{2}\overline{\theta^2}\right)
      + \partial_z \left(\frac{1}{2}\overline{w\theta^2}\right)  =   \beta \overline{w\theta}
        - \tau_{\theta}^{-1}\overline{\theta^2} } \nonumber\\
        & & {} + \frac{1}{2}\partial_z \left(\chi \partial_z \overline{\theta^2}\right) + \frac{1}{2}C_{\theta},   \label{theta2full}
\end{eqnarray}
\begin{eqnarray}
 \lefteqn{\partial_t \left(\overline{w\theta}\right) + \partial_z \left(\overline{w^2\theta}\right)  =  \beta \overline{w^2}
        + (1-\gamma_1) g \alpha_{\rm v} \overline{\theta^2}  
        - \tau_{{\rm p}\theta}^{-1} \overline{w\theta}  } \nonumber\\
        & & {} + \frac{1}{2}\partial_z \left(\chi \partial_z \overline{w\theta}\right)    
        + C_3 + \frac{1}{2}\partial_z \left(\nu \partial_z \overline{w\theta}\right).   \label{wthetafull}
\end{eqnarray}

Here, $K = \overline{q^2/2} = (\overline{u^2 }+ \overline{v^2} + \overline{w^2})$.
We ignore contributions from compressibility ($C_{ii}=0$, $C_{\theta}=0$, $C_3=0$), 
assume a low Prandtl number (drop $\partial_z (\nu \partial_z K)$ and
$\partial_z (\nu \partial_z \overline{w\theta})$) with convection still efficient enough to 
neglect $\partial_z (\chi \partial_z \overline{\theta^2})$ and 
$\partial_z (\chi \partial_z \overline{w\theta})$. In the stationary limit we then obtain
\begin{eqnarray}
   \partial_z \left(\frac{1}{2}\overline{q^2 w} + \overline{p w}\right)  =  g \alpha_{\rm v} \overline{w\theta}
        - \varepsilon,   \label{Kstat}
\end{eqnarray}
\begin{eqnarray}
   \partial_z \left(\frac{1}{2}\overline{w\theta^2}\right)  = 
   \beta \overline{w\theta}
        - \tau_{\theta}^{-1}\overline{\theta^2},    \label{theta2stat}
\end{eqnarray}
\begin{eqnarray}
  \partial_z \left(\overline{w^2\theta}\right)  =  \beta \overline{w^2}
        + (1-\gamma_1) g \alpha_{\rm v} \overline{\theta^2}  
        - \tau_{{\rm p}\theta}^{-1} \overline{w\theta}.    \label{wthetastat}
\end{eqnarray}
In this form the equations also apply to the convective planetary boundary layer 
after taking moist convection into account via $\beta$, the superadiabatic 
or potential temperature gradient.
For these three equations, the {\em non-local transport (flux) terms}
on the left-hand side are {\em exact within the Boussinesq approximation},
as also holds for the terms $g \alpha_{\rm v} \overline{w\theta}$, $\beta \overline{w\theta}$, 
$\beta \overline{w^2}$, and $g \alpha_{\rm v} \overline{\theta^2}$. 

The term $\tau_{\theta}^{-1}\overline{\theta^2}$ results from a closure of
the {\em dissipation rate of (potential) temperature fluctuations}
$\varepsilon_{\theta}$ which  {\em must be positive definite}. Thus, for
further analysis it suffices to consider its closure form. Likewise,
$\varepsilon$ is the {\em dissipation rate of turbulent kinetic energy}, which
is {\em positive definite}, too.

This leaves the closure of $-\Pi_3^{\theta}=-\overline{\theta\partial_z p} =
-\gamma_1 g \alpha_{\rm v} \overline{\theta^2} - \tau_{{\rm p}\theta}^{-1} \overline{w\theta}$ 
as the only non-exact relation one has to pay attention for here. From the Canuto-Dubovikov 
turbulence model \cite{canuto96d,canuto97e} one obtains $\gamma_1=1/3$  
(values between 1/3 and 1/5 are suggested in the literature) whereas the time scale 
$\tau_{{\rm p}\theta}$, just as $\tau_{\theta}$, is necessarily positive (see also \cite{canuto98b}). 
Hence, this closure takes into account that $\Pi_3^{\theta}$ in principle changes sign
like $\overline{w\theta}$ although not necessarily at the same location (the
exact shift being subject to closure uncertainty). As is shown
below this is not relevant to recover the countergradient flux nor is
it the key element to obtain a negative buoyancy (convective, enthalpy) flux,
even though it is tempting to conclude just that from a purely local point of view.
The latter clearly leads to wrong conclusions on overshooting.

Consider Eq.~(\ref{Kstat}) and abbreviate $J \equiv \overline{w\theta}$. 
For the non-local pressure flux $\overline{p w}$ we note that it is often closed
by assuming it to be a fraction of the flux of kinetic energy and thus being proportional
to $\overline{q^2 w}$. It is in any case a non-local term and hence does not change
any of the following conclusions where for simplicity we assume it to be absorbed into 
the term $\propto \overline{q^2 w}$.

Assume $J > 0$. Then, non-zero velocities require $\varepsilon > 0$ and we 
can find a local solution, $\frac{1}{2}\overline{q^2 w}=0$, inside a convectively 
unstable zone. This is essential for MLT to work!

Assume $J < 0$, then no non-trivial local solution exists. Thus, we must have 
$\partial_z(\frac{1}{2}\overline{q^2 w})\neq 0$ for any solution with non-zero velocities.
This realization is at the heart of all ``1-equation models'' of overshooting
which consider a {\em dynamical equation for turbulent kinetic energy or its
flux}, whether as a differential equation, in an integral form (I.W.~Roxburgh's
integral constraint, \cite{roxburgh78b} and \cite{roxburgh89b}), or in 
a formalism based on that ingredient, among others (such as the plume model 
for convective penetration by J.-P.~Zahn, \cite{zahn91b}).

However, this structure of a convective overshooting zone is {\em at variance 
with observations from meteorology}. Following their publication \cite{priestley47b} in 1947, 
the countergradient transport of heat in atmospheric turbulent flows became known as the 
Priestley-Swinbank effect. Its first convincing explanation  was given in 1966 by 
J.W.~Deardorff \cite{deardorff66b} who made it clear why there is a region with 
a {\em potential temperature gradient opposite to the direction of the turbulent heat flux} 
by analyzing a variant of Eq.~(\ref{theta2stat}) with data from meteorology and laboratory 
experiments (work with G.E.~Willis 1966, see \cite{deardorff66b} --- an earlier suggestion
by Deardorff \cite{deardorff61b} had still lacked the observational data to confirm the idea 
further elaborated in \cite{deardorff66b}). 

Consider $\beta > 0$ and $J > 0$ in Eq.~(\ref{theta2stat}). This permits local solutions, where
$\frac{1}{2}\overline{w\theta^2}=0$.

Now if $\beta < 0$ while $J > 0$ there can be no non-trivial local solution. Instead,
$\partial_z(\frac{1}{2}\overline{w\theta^2}) \neq 0$. For this reason, {\em $\overline{w\theta^2}\neq const.$
is the key ingredient for the existence of the countergradient region in overshooting}.

If both $\beta < 0$ and $J < 0$, Eq.~(\ref{theta2stat}) would permit a local,
non-trivial solution, but this is not the case for Eq.~(\ref{Kstat}) for which already $J < 0$ excludes
a non-zero local solution. Hence, we {\em need to model both $\partial_z(\frac{1}{2}\overline{q^2 w})$ 
and $\partial_z(\frac{1}{2}\overline{w\theta^2})$ by dynamical equations} or closure conditions.

Deardorff \cite{deardorff66b} also realized the connection between  $\overline{w\theta^2}$ and
$S_{\theta}=\overline{\theta^3}/(\overline{\theta^2})^{3/2}$. He noticed that skewness and the location 
of convective driving are related to each other (heating from below or both heating from below and 
cooling from above as in the water tank experiment with Willis). This was corroborated by work reported, 
among others, in \cite{wyngaard87b,moeng89b,moeng90b,holtslag91b,wyngaard91b,schumann93b,piper95b},
more than 20 years after Deardorff's pioneering work, for then numerical simulations could be used to support 
Reynolds stress budgets and plume model based arguments.

In the absence of a non-local flux $\overline{w^2\theta}$, the specific form
of Eq.~(\ref{wthetastat}) is compatible with a solution where $\beta > 0$ and $J > 0$. 
This is also the case for $\beta < 0$ and $J > 0$ and even for $\beta < 0$ and $J < 0$. 
Hence, Eq.~(\ref{wthetastat}) appears to permit local solutions for convectively stable
and unstable stratifications. However, whether the correct signs of positive definite quantities 
are realizable is only a necessary condition for local solutions to exist, not a sufficient one.
In this sense, only a detailed analysis of $\Pi_3^{\theta}$ itself can demonstrate the necessity 
for a non-zero $\partial_z(\overline{w^2 \theta})$. Modern LES and observational data of the PBL 
show that this is indeed the case (\cite{canuto94b},  \cite{mironov00b}, \cite{gryanik02b}):
also this term is clearly non-zero in overshooting regions.

This chain of arguments also holds for the compressible case since using the density weighted
Favre average \cite{favre69b} as in \cite{canuto97b} leads to structurally similar equations
and the smallness of additional terms can be established from detailed budget equations
(unclosed Reynolds stress equations with contributions computed from 3D numerical simulations,
as in \cite{mironov00b} for the Boussinesq case).

A {\em 3-equation model} (for $K$, $J$, and $\overline{\theta^2}$) such as that one by 
R.~Kuhfu{\ss}, \cite{kuhfuss86b}, appears to be needed to capture the behaviour of 
$\beta$ and $J$ in overshooting zones. To also self-consistently account for how the ratio between
vertical and total kinetic energy changes with depth an equation for $\overline{w^2}$ has to be
added, too, as in the Reynolds stress models of D.R.~Xiong and V.M.~Canuto 
(\cite{xiong86b,xiong97b},\cite{canuto11b}). Computationally, such non-local models are much 
more costly than the ``recipies'' used for convective overshooting in stellar evolution calculations.

Finally, overshooting is a {\em fundamentally non-local} phenomenon, a result demonstrated 
in meteorology more than 50 years ago without requiring any numerical simulations even though
the latter are useful to further corroborate this finding.

\subsection{Modelling non-locality} \label{sec-5.2}

One important tool to test convection models in general and closure assumptions  
in particular are budget equations derived from unclosed ensemble averages. Assuming
again a quasi-ergodicity hypothesis one can calculate these budgets from
numerical simulations. Examples for the case of the planetary boundary layer of
the Earth can be found, for instance, in \cite{mironov00b} who also computed these
statistics for two different rotation rates in addition to the non-rotating case. Indeed,
the flux of temperature fluctuations, $\overline{w\theta^2}$, is shown to be essential
for a balanced, quasi-stationary stratification for both cases of rotation and for the
non-rotating scenario. The same kind of simulation can also be used to probe closure
expressions (\cite{zilitinkevich99b,canuto01b}). For a most convincing test the
simulation data can also be combined with observational data, as has been done
by \cite{gryanik02b,cheng05b} for the tests of their closure models.

Claiming universality of a closure suggested, say, for $\overline{w\theta^2}$ or
$\overline{w^2\theta}$, however, requires a larger region of applicability.
For instance, in \cite{kupka17b} it was demonstrated that the two-scale
mass flux closure Eq.~(\ref{eq.wt2}) yields excellent results both
in comparison with a numerical simulation of solar granulation and with a numerical 
simulation of convection in a moderately hot ($T_{\rm eff} \sim 11800\,\rm K$)
DA type white dwarf. It was also shown that compressibility (as accounted for
by using Favre averages) has only very little impact on this result. The traditional
one-scale mass flux closure of this quantity, which ties it to the skewness of
the vertical velocity instead of the temperature field, fails in the overshooting
regions as it violates a sign symmetry. Thus, in spite of its complexity,
the non-local flux of temperature fluctuations can be modelled fairly well
over a very large parameter space by a formally rather simple algebraic 
expression. This is not the case for all closure relations, but it does hold for
quite a few (see \cite{kupka07f} for some further examples).

\subsection{The DA white dwarfs} \label{sec-5.3}

White dwarfs turn out to be particularly useful objects for the study of convective 
overshooting. At effective temperatures around 12000~K, DA white dwarfs are 
characterized by a thin atmosphere which enshrouds a zone of notable temperature 
change which in turn contains an isothermal, electron degenerate core. Around 
$T_{\rm eff} \sim 12000\,\rm K$ the surface convection zone, which is caused by 
(partial) ionization of hydrogen, is thin enough so as to be barely deeper than the 
atmosphere itself. Thus, a box-in-a-star-type simulation in Cartesian geometry is 
possible: for a box with a depth of about 8~km and a width around 15~km, the 
curvature of a star with a radius of 6000~km is negligible near its surface. 

The DA white dwarfs have originally been characterized by their hydrogen lines
in the optical spectrum and the lack of helium and metal lines in the same 
wavelength region. Extensive spectroscopic observations have demonstrated
that about 25\% of these objects in fact do have metal lines 
(DAZ stars, \cite{zuckerman03b,gianninas14b,koester14b}).

The standard model on how white dwarfs could assemble a metal polluted
atmosphere is steady-state accretion, as discussed in \cite{koester06b,koester09b},
who provide both the theoretical models and further data needed for such calculations.
Possible accretion sources are planetary systems and remnants of them (some authors
have previously doubted this type of source, but it has now become the standard explanation).
However, diffusion inside the stars spreads the accreted material into the star. 
Counteractions of gravitational setting and radiative diffusion on a microscopic scale
vary element by element. In addition, a convection zone at the stellar surface provides
mixing on very short time scales, practically instantaneously. This defines the size of the 
reservoir into which accretion occurs and is counteracted by diffusion. Overshooting
extends the convectively mixed region. But by how much and how reliable are models
of this process? Clearly, numerical simulations are needed to quantify this more accurately.

\subsection{3D simulations of overshooting} \label{sec-5.4}

In numerical simulations of overshooting the following length and time scales have 
to be considered either directly or indirectly (see \cite{kupka17b,kupka18b}). 
They are first of all based on the underlying physical processes.

Among the length scales of interest are:
\begin{itemize}
\item The length scale of the maximum of kinetic energy transport which 
        is a function of radius, $H(r)$. At the stellar surface this is roughly
        the horizontal extent of a granule: $H(R) \sim H_{\rm gran}$.
\item The length scale of viscous dissipation $l_d$, also known as the Kolmogorov scale.
        In numerical simulations of stellar convection with realistic microphysics this scale 
        is of course always unresolved and a much larger turbulent viscosity is taking over
        its role. This shifts dissipation of kinetic energy to much larger length scales.    
\item The size of the radiative (or thermal) boundary layer, $L_t$. Once more a function 
        of radius it is hence different for the upper and for the lower boundary
        of a convection zone. An estimate (\cite{kupka17b}) is given by
        $L_t \gtrsim \delta \sim {\rm Pr}^{-1/2}\, l_d$. Here, $\delta$ is the thermal boundary
        due to (radiative) diffusion. The Prandtl number compares momentum to (radiative) heat
        diffusion, ${\rm Pr} = \nu/\chi$. The estimate can be improved, \cite{zahn91b}, which
        does not change the conclusions drawn below.
\end{itemize}

The most important time scales are:
\begin{itemize}
\item The thermal relaxation time scale $t_{\rm therm}$, which for non-degenerate
        stars can be approximated by $t_{\rm therm}(x) \sim t_{\rm KH}(x)$, as discussed
        in detail in \cite{kupka17b} (see \cite{kippenhahn94b}). Both time scales are
        again defined in a more general form as functions of location and the latter is given by
        \begin{equation}  \label{eq.def_t_kh}
            t_{\rm KH} = \left(-3\int_{M_s(r_{\rm a})}^{M_s(r_{\rm b})} p\, \rho^{-1}\,{\rm d}M_s\right) / {\cal L}.
        \end{equation}
        Here, $M_s(r)$ is the mass contained from the surface of a star till a radius $r$ and
        $r_b > r_a$, while ${\cal L}$ is the luminosity (at the local radius or constant).
        $p$ and $\rho$ are pressure and density.
\item The convective turn over time $t_{\rm conv}$,
         \begin{equation}  \label{eq.def_t_conv}
             t_{\mathrm{conv}} =\int_{r_{\rm a}}^{r_{\rm b}} u_x^{-1}(r)\,{\rm d}r.
         \end{equation}
         Here, $u_x^{-1}(r)$ is the ensemble average of the root mean square difference
         between the local vertical velocity and its horizontal average. 
\item This allows the definition of the relaxation time $t_{\rm rel}$ (see Sect.~\ref{sec-4.3} and \cite{kupka17b}) 
         as follows:
         \begin{equation}  \label{eq:trel}
             t_{\mathrm{rel}} \approx \max(t_{\mathrm{conv}}(r_{\rm bottom}),t_{\mathrm{therm}}(r_{\rm rel})).
         \end{equation}
         Thus, relaxation has to occur both with respect to the kinetic energy of the flow
         ($t_{\mathrm{conv}}$) and the potential (thermal) energy of the stratification ($t_{\rm KH}$)
         and hopefully, $r_{\rm rel} > r_{\rm bottom}$.
\end{itemize} 
These scales are linked to crucial dimensionless ratios: the Reynolds number ${\rm Re}=(u(l)l)/\nu$, 
the P\'eclet number ${\rm Pe}={\rm Re\,Pr}= (u(l)l)/\chi$, and the Rayleigh number
${\rm Ra}=(t_{\rm visc}\,t_{\rm rad})/t_{\rm bouy}^2$. Evidently, these are {\em scale dependent} numbers and the 
length scale they are based on is crucial for their interpretation. Thus, $u(l)$ refers to a vertical (or total) velocity 
which can relate either to a length scale on which most kinetic energy is transported, $u(H(r))$, or to a scale 
$l$ on which only some small scale fluctuations occur. Note that $t_{\rm visc} = l^2/\nu$ and $t_{\rm rad} = l^2/\chi$ 
(radiative diffusion) and $t_{\rm bouy} = N_{\rm BV}^{-1}$, the inverse of the Brunt-V\"ais\"al\"a frequency.

The scales have to be accounted for in 2D and 3D numerical simulations of overshooting. To this end 
simulation based (effective) dimensionless parameters can be introduced (cf.\ \cite{kupka09b,kupka17b}).
For the 3D case one can estimate whether it is possible to resolve turbulence generated by shear between
resolved upflows and downflows by estimating ${\rm Re_{eff}} \sim (H/h)^{4/3}$. Here, $h$ refers to the grid spacing
used and $H$ to the scale to which the {\em effective Reynolds number} refers to. For judging whether the
shear between granules and downflows can generate turbulence, $H$ is set to $H_{\rm gran}$. Vortex tubes
may be found at rather moderate values of a few 100 or less, but their character changes once $Re_{\rm eff}$
significantly exceeds values of 2000. To check this statement it is instructive to compare some simulations
published in the literature: large vortex tubes are clearly visible in downflow lanes of Fig.~13 and Fig.~14 of \cite{stein00b}
who at $h \approx 23.7\,\rm km$ horizontal resolution and assuming $H_{\rm gran} \approx 1300\,\rm km$ achieve 
${\rm Re_{eff}}  \approx 208$. Somewhat more small turbulence and a highly upwards reaching ``tornado'' are visible in Fig.~15 
and 16 of \cite{muthsam10b}, where $h \approx 9.8\,\rm km$, whence ${\rm Re_{eff}}  \approx 677$. A change only occurs at 
even higher ${\rm Re_{eff}}$ and was first shown in \cite{muthsam11b} who demonstrated the transition to a highly turbulent flow once 
the resolution is increased from 7.4~km (and thus ${\rm Re_{eff}}  \approx 984$) to 3.7~km (with ${\rm Re_{eff}}  \approx 2479$)
and can be inspected by comparing Fig.~\ref{fk.fig-1} with Fig.~\ref{fk.fig-2} in Sect.~\ref{sec-3} further above.
This intense level of turbulence is also observed for a more quiet state of granule evolution that is shown
in Fig.~13 of \cite{kupka17b} (where $h \approx 2.5\,\rm km$ and thus ${\rm Re_{eff}}  \approx 4182$). Thus, only once 
${\rm Re_{eff}} $ clearly exceeds 1000, small scale turbulence can be observed on the computational grid itself. 
Many simulations of astrophysics in general and of stellar convection in particular have ${\rm Re_{eff}}  \lesssim 1000$.
They feature rather thick vortex tubes and do not credibly show a Kolmogorov inertial range: at lower resolution 
the kinetic energy carrying scales and the energy dissipating grid scale are still close and thus tightly coupled to each 
other and the hypotheses underlying the Kolmogorov theory cannot hold (cf.\ \cite{pope00b,tsinober09b}). Note that 
the {\em box Reynolds number}, ${\rm Re_{box}} = (L/h)^{4/3}$, is not the best guide to judge 
on the level of turbulence: $L$ may include stable layers, get boosted by horizontal box extent, or even $L \gg H_{\rm gran}$.

A second important parameter is the {\em effective Prandtl number}, ${\rm Pr_{eff}}$. It compares turbulent or grid viscosity with
radiative or microscopic heat diffusivity. In numerical simulations of convection in astrophysics one usually aims at achieving 
$1 \gtrsim {\rm Pr_{eff}} \gtrsim 0.1$. Another important parameter compares convective to conductive heat transport on the grid: 
the effective P\'eclet number, ${\rm Pe_{eff}} = {\rm Re_{eff}}\, {\rm Pr_{eff}}$. In astrophysical applications that fulfil 
${\rm Pr_{eff}} \lesssim 1$ it is possible to achieve ${\rm Pe_{eff}} \lesssim 1000$. It may be tempting to invoke
2D simulations to boost ${\rm Pe_{eff}}$, since they follow a different scaling (see \cite{lesieur08b}): ${\rm Re_{eff}} \sim (H/h)^2$, 
whence ${\rm Pe_{eff}} \lesssim 10^5$.  However, this scaling is bought at the price of an inversely operating energy cascade
and 2D flows are much more often non-ergodic, \cite{tsinober09b}. Estimates of overshooting based on 2D simulations
thus have to be done with great care.

What happens if some of those scales are ignored in a numerical simulation of overshooting? 
If $t_{\rm rel} \ll \max(t_{\mathrm{conv}}(r_{\rm bottom}),t_{\mathrm{therm}}(r_{\rm rel}))$, then the
temperature gradient $dT/dr$ and the extent of overshooting may be unrealistic, since the relaxed 
structure may not have been known in advance. If ${\rm Pe_{eff} \ll Pe}$, we have to expect the
amount of overshooting to become wrong, too, since the thermal structure and the amount of
overshooting are highly sensitive to $\rm Pe$ as was shown in \cite{brummell02b}. The role of $\rm Pe$
and the case  ${\rm Pr_{eff}} > 1$ are considered further in Sect.~\ref{sec-5.5}. If, finally,
$L_t$ or $\delta$ remain unresolved, one might have to deal with flow artifacts, as are known
from numerical simulations of Rayleigh-B\'enard convection in similar situations.

\subsection{The role of the P\'eclet number} \label{sec-5.5}

How do numerical simulations of the solar surface, of the solar tachocline, and of shallow 
convection zones in DA white dwarfs differ from each other? 

For the Sun, at the surface $H \sim H_{\rm gran} \sim 1300\,\rm km$, $l_d \sim 1\,\rm m$, $L_t \sim 30\,\rm km$,
${\rm Pr} \sim 10^{-9}$ and thus ${\rm Pe} \sim 10$ (further details on how to estimate those numbers are given in \cite{kupka17b}).
The requirements ${\rm Pe_{eff}} \lesssim 1000$ and $1 \gtrsim {\rm Pr_{eff}} \gtrsim 0.1$ can easily be fulfilled and highly realistic
simulations are possible for those layers. Indeed, for a simulation of solar granulation within a box that is 4~Mm deep and convective 
for all layers more than 0.7~Mm away from the top and which is quasi-adiabatic for layers that are more than
1.5~Mm away from the top, at $h=11\,\rm km$, one finds $t_{\rm rel} = \max(t_{\rm KH}(1.5\,{\rm Mm}),t_{\rm conv}(4\,{\rm Mm})) 
= \max(45\,{\rm min}, 57\,{\rm min})	< 1\,{\rm hr}$. At a time step $\Delta t = 0.25\,\rm s$ this requires $N_t \sim 1.4\times 10^4$
time steps. This is easily affordable and in fact is a problem with a computational complexity $C < 1$ (see Sect.~\ref{sec-4.6}).

The problem of simulating overshooting underneath the solar convection zone is of a very different
kind. Taking again the values estimated in \cite{kupka17b}, one finds $l_d  \sim 1\,\rm cm$, $L_t \sim 1\,\rm km$ as
a more realistic estimate in comparison to $\delta \sim 30\,\rm m$, ${\rm Pr} \sim 10^{-7}$, and ${\rm Pe} \sim 10^6$
for the bottom of the solar convection zone. If we want to keep $1 \gtrsim {\rm Pr_{eff}} \gtrsim 0.1$, then necessarily 
${\rm Pe_{eff}} \lesssim 1000$ and the resulting overshooting calculation cannot be expected to be realistic. 

What happens, if one gives up on $1 \gtrsim {\rm Pr_{eff}} \gtrsim 0.1$ to achieve a higher value of ${\rm Pe_{eff}}$? 
To answer this question it is helpful to recall some discussion originally presented in Sect.~4.1 of \cite{kupka09b}
and elaborate it with some additional comments. A stable numerical simulation requires that the grid cell Reynolds number
${\rm Re_{grid}} = U(h) h / \nu_{\rm eff} = O(1)$. In this context $\nu_{\rm eff}$ is the viscosity the numerical scheme 
has to provide to achieve ${\rm Re_{grid}} = O(1)$. For the linear advection equation and basic finite difference schemes 
one can prove that ${\rm Re_{grid}} \leqslant 2$ (see Chap.~6.4 in \cite{strikwerda89b}). For that case $U(l)=a=\rm const.$
at all $l$ and thus ${\rm Re_{eff}} = U(l)\,l / \nu_{\rm eff} \ll {\rm Re} = U(l)\,l / \nu$ if $\nu_{\rm eff} \gg \nu$. One can rewrite
this into ${\rm Re_{eff}} = U(l)\,(l/h)\,h / \nu_{\rm eff} =(l/h)\,(U(l)\,h/\nu_{\rm eff}) \sim l/h$, since for a stable simulation
$U(l) \,h / \nu_{\rm eff} = U(h) \,h / \nu_{\rm eff} \sim O(1)$ and $U(l)=U(h)$ here.

For a turbulent convective flow this is too pessimistic because $U$ in the definition of ${\rm Re_{grid}}$ refers to 
$U(h)$ at the grid cell level. Contrary to linear advection, however, $U(h)$ is smaller in magnitude than $U(l)$ at the 
energy carrying scales. Thus, $U(l)\,l/\nu_{\rm eff} \sim O(1)$ would require an unnecessarily viscous simulation 
whereas $U(l)\,l/\nu_{\rm eff} \gg 1$ is both desirable and possible. Thus, one can insert $U(h)/U(h)$ in the previous 
expression for ${\rm Re_{eff}}$ and rewrite it into ${\rm Re_{eff}} = (l/h)\,(U(l)/U(h))\,(U(h)\,h/\nu_{\rm eff})$ from which 
we see that it is sufficient to demand $(U(h)\,h/ \nu_{\rm eff}) = {\rm Re_{grid}} = O(1)$ for a stable simulation. Violating 
the latter constraint though would cause unwanted oscillations in the numerical solution.

It is now left to explain the estimate $(l/h)\,(U(l)/U(h)) \sim (l/h)^{4/3}$ to justify the definition ${\rm Re_{eff}} \sim (H/h)^{4/3}$
as the effective Reynolds number of a stable, 3D numerical simulation of a turbulent flow. To this end the first and
second hypotheses of Kolmogorov have to be invoked (cf.\ \cite{pope00b} for further details about them). The first hypothesis 
assumes that the velocity difference of two fluid elements separated by a distance $\lambda$ depends only on the flux of kinetic 
energy between different scales around $\lambda$ and on the viscosity $\nu$. Clearly, $l > \lambda$. In practice, 
also $l \gg \lambda$ can be necessary. The second hypothesis assumes that there is a restricted range called 
inertial range where the velocity difference is also independent of $\nu$. Clearly, $\lambda > l_d$, where $l_d$ is the 
Kolmogorov scale for which $U(l_d)\,l_d/\nu \sim O(1)$. For the inertial range the energy flux is proportional to 
$u(\lambda)^3/\lambda$ and independent of $\lambda$ (see \cite{pope00b}). Consequently,
$U(l)/U(h) \sim (l/h)^{1/3}$ and $(l/h) \,(U(l)/U(h)) \sim (l/h)^{4/3}$ which completes the derivation. Note that it has been
important to select the ``correct'' value of $U$ from a scaling relation that had to be derived separately (in this case from 
Kolmogorov's theory). For this reason, in 3D one finds ${\rm Re_{eff}} \sim (H/h)^{4/3}$ whereas in 2D, ${\rm Re_{eff}} \sim (H/h)^2$.
Although for low numerical resolution the Kolmogorov inertial range scaling cannot hold, the kinetic energy of 
such a simulation varies slowly and smoothly around the energy carrying scales and decreases quite steeply close
to the grid scale (this is also a result of the construction principles behind numerical and artificial viscosities). This allows
an $O(1)$ estimate from ${\rm Re_{eff}} \sim (H/h)^{4/3}$ for those cases.

Having clarified the physical meaning of ${\rm Re_{eff}}$ it is now possible to clarify the role of ${\rm Pr_{eff}}$ and 
${\rm Pe_{eff}}$. Their definitions are straightforward following their molecular counterparts, \cite{kupka09b}:  
${\rm Pr_{eff}} = \nu_{\rm eff} / \chi$ and ${\rm Pe_{eff}} = {\rm Re_{eff}} \, {\rm Pr_{eff}}$. There is a pitfall ahead. At face 
value, $\nu_{\rm eff}$ cancels in this definition and ${\rm Pe_{eff}} = {\rm Pe}$ even if the simulation has $\nu_{\rm eff} \gg \nu$.

Following again \cite{strikwerda89b} it turns out that also for the grid P\'eclet number ${\rm Pe_{grid}} = U \, h / \chi_{\rm eff} = O(1)$.
More precisely, ${\rm Pe_{grid}} \leqslant 2$ for linear heat diffusion with advection, in addition to ${\rm Re_{grid}} \leqslant 2$. 
But that condition also has to be fulfilled for the hydrodynamical equations in the sense of ${\rm Pe_{grid}} = O(1)$ alongside
${\rm Re_{grid}} = O(1)$, otherwise the positivity of diffusion is violated and fluctuations of energy pile up at the grid scale
(this is revealed by a more detailed analysis of Eq.~(\ref{sec-2-eq_energy-std}) which describes how the advection of internal
energy and radiative diffusion contribute to energy transport).
Assume now that $\nu_{\rm eff} < \chi$ or ${\rm Pr_{eff}} < 1$. Then it is sufficient to set $\chi_{\rm eff} = \chi$ and both
conditions on ${\rm Re_{grid}}$ and ${\rm Pe_{grid}}$ can be fulfilled at the same time since $\nu_{\rm eff}$ was assumed 
to be large enough to make ${\rm Re_{grid}} = O(1)$ in first place. It thus makes perfect sense to identify 
${\rm Pe}={\rm Pe_{eff}}$. This is the case for simulations of convection at the solar surface and for the DA white dwarf we
consider below. 

But in other cases $\nu_{\rm eff} \gg \chi$, i.e.\ ${\rm Pr_{eff}} \gg 1$, may be inevitable when assuring that ${\rm Re_{grid}} = O(1)$. 
In this case $U \, h / \chi \gg O(1)$ by the definition of ${\rm Re_{grid}}$. Thus, the simulation would be 
unstable unless there is a sufficiently large numerical heat diffusion $\chi_{\rm eff} \gg \chi$ which permits 
${\rm Pe_{grid}} = U \, h / \chi_{\rm eff} = O(1)$. In subgrid scale models of turbulent flows this is expressed by setting
a {\em turbulent Prandtl number}, ${\rm Pr_{turb}} = \nu_{\rm eff} / \chi_{\rm eff}$ which is typically somewhat less than 1.
Then again the stability conditions are fulfilled. However, such a change also requires to {\rm redefine} 
${\rm Pe_{eff}} = U(l)\,l / \chi_{\rm eff} = (U(l)\,l / \nu_{\rm eff}) (\nu_{\rm eff}/\chi_{\rm eff}) = {\rm Re_{eff}} \, {\rm Pr_{turb}}$.

For a numerical simulation where $1 \gtrsim {\rm Pr_{eff}} \gtrsim 0.1$ one has ${\rm Pr_{turb}} \sim {\rm Pr_{eff}}$ and one
achieves ${\rm Pe} = {\rm Pe}_{\rm eff}$. If instead the resolution of the simulation is so low that ${\rm Pr_{eff}} > 1$ or even
${\rm Pr_{eff}} \gg 1$, then the simulation can only be stable, if either explicitly through a turbulent heat diffusivity or
by some other form of numerical viscosity associated with the spatial discretization, or implicitly, through a damping
included in the time integration scheme, sufficient {\em numerical heat diffusion} is introduced, i.e., $\chi_{\rm eff} > \chi$
or even $\chi_{\rm eff} \gg \chi$. In this case one has to use the second definition of ${\rm Pe_{eff}}$, ${\rm Pe_{eff}} = 
{\rm Re_{eff}} \, {\rm Pr_{turb}} <  {\rm Re_{eff}}$. Thus, ${\rm Pe_{eff}} \lesssim {\rm Pe}$ or even  ${\rm Pe_{eff}} \ll {\rm Pe}$.

In conclusion, if the grid resolution is too coarse to resolve (radiative) heat diffusion, then the P\'eclet number which
can be achieved by a numerical simulation using such a grid is limited basically by the effective Reynolds number. 
This is just what happens when performing a numerical simulation for overshooting below the solar convection zone:
the P\'eclet number which can be achieved by the simulation is several orders of magnitudes smaller than the 
actual P\'eclet number of the problem. Thus, systematic differences between overshooting predicted from such simulations 
and the actual overshoot have to be expected.

Since ${\rm Pe} \sim 10^6$ at the bottom of the solar convection zone, realistic P\'eclet numbers cannot be 
reached for currently affordable spatial resolutions in simulations of global solar convection. Their relaxation is
also troublesome. With $t_{\rm rel} = \max(t_{\rm KH}(180\,{\rm Mm}),t_{\rm conv}(180\,{\rm Mm})) = 
\max(\sim 10^6\,\rm yrs, 1\,month)$ we obtain due to the problem of long thermal relaxation and considering time 
steps from 0.25~sec to a few hours that $N_t \sim 2 \times 10^9 \dots 10^{14}$. This problem has already
been addressed in Sect.~\ref{sec-4.6}.

Overshooting below stellar convection zones is more easily studied for realistic microphysics when remaining closer to the 
surface of a star. For DA white dwarfs it is possible to consider cases with strong convection that are readily affordable with
current computational resources. The object studied in \cite{kupka18b} is a DA white dwarf with $T_{\rm eff} = 11800\,\rm K$, 
a surface gravity $\log(g)=8$, and pure hydrogen composition. The simulation box was designed for a layer about 
7.5~km deep for which the thermal structure deviates from a radiative one for about the top 4~km. In this case, 
$H \sim H_{\rm gran} \approx \rm 0.45\,km \dots 1.35\,km$. Likewise, $l_d \sim 25\,\rm cm$ $\delta \sim 3.4\,\rm km$,
${\rm Pr} \sim 5.4 \times 10^{-9}$. Consequently, ${\rm Pe} \sim 10 \dots 30$. With $h \approx 30\,\rm m$ it is possible
to achieve ${\rm Pr_{eff}} \lesssim 0.2$ while ${\rm Re_{eff}} \sim 40 \dots 170$. Consequently, ${\rm Pe_{eff}} \sim 10 \dots 30$
and thus ${\rm Pe_{eff}} \sim {\rm Pe}$. This permits realistic 3D hydrodynamical simulations of overshooting and indeed the
entire range of ${\rm Pe} \lesssim 1$ to ${\rm Pe} \gtrsim \mbox{\rm a few} 100$ is accessible for $T_{\rm eff}$ in the 
range of 14000~K to 11600~K at the same values of surface gravity and chemical composition. The realistic upper boundary 
condition of this case also prevents artifacts and can hence yield reliable benchmarks for non-local convection models.
Thus, $t_{\rm rel} = \max(t_{\rm KH}(4\,{\rm km}),t_{\rm conv}(\sim 7\,{\rm km})) = \max(25\,{\rm s}, 30\,{\rm s})$.
With a relaxation time of 30~s at a time step $\Delta t = 22\,\rm ms$ we have $N_t \sim 1.4\times 10^5$. Hence, 
a relaxed simulation resolving the relevant scales in space and time has been possible.

\subsection{Previous research and a deep DA WD simulation} \label{sec-5.6}

DA type white dwarfs have raised interest as goals of numerical studies of convection
due to a number of specific astrophysical questions related to pulsational stability,
spectroscopic determinations of stellar parameters, and the already introduced 
problem of mixing of accreted material (Sect.~\ref{sec-5.3}). Another, practical
advantage are the readily fulfillable demands on resolution and on statistical relaxation 
(Sect.~\ref{sec-5.5}). DA white dwarfs have hence been studied with numerical simulations
for quite a while. The earliest work considered 2D simulations, \cite{freytag96b},
which have demonstrated the existence of strong overshooting below the actually
convectively unstable region in hotter objects ($T_{\rm eff}$ of 12200~K, for example)
and indicated an exponential decay of the velocity field in that region. While those findings
appeared convincing for higher values of effective temperature ($T_{\rm eff} \approx 13400\,\rm K$),
a need for deeper simulation boxes can also be concluded from the velocity profiles shown
for lower $T_{\rm eff}$ in their work. This turned out to be at variance with velocity
profiles derived from Reynolds stress models, \cite{montgomery04b}. There, in the 
overshooting region proper where notable negative convective fluxes exist (plume-dominated
region) and in the lowest part of the countergradient region (as described by Deardorff),
the velocity field was found to decay linearly as a function of distance from the convectively
unstable zone. Similar was found in the equivalent region of DB white dwarfs studied
in \cite{montgomery04b} and for A-type main sequence stars analyzed in \cite{kupka02b}.

Eventually, in \cite{tremblay11b,tremblay13b,tremblay15b} 3D numerical simulations of surface
convection in DA white dwarfs have been presented and analyzed in \cite{tremblay15b} with
respect to overshooting. Their results appeared to confirm the previous findings of \cite{freytag96b}.
However, closer inspection of their Fig.~4 and Fig.~5 revealed differences in the profiles
of convective flux and vertical root mean square velocity between hotter ($T_{\rm eff} \gtrsim 13000\,\rm K$) 
and colder ($T_{\rm eff} \lesssim 12500\,\rm K$) models when compared as a function of distance from the 
convectively unstable region. Since indications for the influence of the lower vertical simulation boundary
could be argued for (their Fig.~12) in case of the cooler models due to their larger region of convective
instability and overshooting, a detailed simulation for this parameter region for a larger volume 
appeared necessary. This has been presented in \cite{kupka18b}. There, the authors identified
the presence of self-excited waves dominating the lower part of the simulation box and thus
also the overshooting region at greater distance from the convectively unstable zone. Closer
to the latter, agreement was found with the Reynolds stress model discussed in \cite{montgomery04b}:
a linear decay of the velocity field provides clearly a better fit for the plume-dominated region
of overshooting than an exponential model. The latter only begins to work when overshooting
finally fades away, in the transition layer to the wave-dominated region and within the latter.
A detailed analysis of the statistical properties of the different regions of the overshooting zone
was given and a lower estimate for the convectively mixed region was suggested.

\subsection{Conclusions} \label{sec-5.7}

The modelling and the quantitative description of overshooting due to
stellar convection zones allows some interesting conclusions drawn when
considering a case for which physically reliable numerical simulations 
can be made. This is possible for the case of DA white dwarfs. Lessons
learned from that work include the following.

\begin{itemize}
\item Concerning simulation extent and resolution a sufficiently wide (high aspect ratio) and well resolved 
        numerical simulation is necessary to obtain accurate statistics. For the case considered
        in \cite{kupka18b} it must extend to more than 4 pressure scale heights below Schwarzschild unstable zone 
        to obtain a wave-dominated region which is not strongly influenced by the lower boundary condition
        (this also depends on the stellar parameters considered).
\item With respect to the properties of the overshooting zone it is interesting to note that there is a
        linear decay of root mean squared velocities in the plume-dominated region just as in the Reynolds stress models
        whereas an exponential decay can be argued for only in the wave-dominated region and in the transition region
        to the latter. This exponential decay appears to be even faster for horizontal velocities.
\item A conservative estimate for accretion concludes a 1.6 to 2.5 times higher accretion rate is needed due to the larger
        amount of mass mixed by convective overshooting compared to the convectively unstable zone alone. This value is 
        obtained if only the plume-dominated region is considered. It increases to a factor of 3.2 to 6.3, if the linear extrapolation
        suggested in \cite{kupka18b} is used. 
\item The main limitations of this approach are the strong modes excited in the simulation which are enhanced by
        the small mode mass of the simulation box. The mixing caused by waves requires a proper scaling of amplitudes
        since otherwise very strong overestimations of its efficiency can occur.
\end{itemize}      

\newpage

\section{Summary}  \label{sec-6}

Given the necessarily limited extent of the original lectures the associated notes presented in this
review have focussed on some core topics of the physics of thermal convection and its study in 
astrophysics. A guide to further literature is hence given in the introductory Sect.~\ref{intro}
where readers can also find reviews on subjects other than those covered here. The physics of convection
is introduced in Sect.~\ref{sec-2} first from a heuristic point of view and explained further by examples 
from astrophysics and geophysics. Some of the main implications of this process for astrophysics are then 
presented. The section concludes by first repeating the classical linear stability analysis which yields
the Schwarzschild criterion of convective instability, followed by introducing the basic hydrodynamical
conservation laws, on which this analysis is actually based on. A first hint on problems ahead is then given
by introducing the challenge posed by the huge spread of scales caused by convection in stellar scenarios.
The different modelling approaches to convection are introduced in Sect.~\ref{sec-3}. Emphasis is given
on modern approaches including non-local models of different types, advanced closure models, and by
a comparison of these methods to those used in hydrodynamical simulations. The latter are then
discussed in Sect.~\ref{sec-4}. Since there are several recent reviews on this subject, the focus of
this part is on important basic questions which nevertheless find less attention: the relation of simulations
to statistical physics, the uniqueness of numerical solutions which entails the inevitable introduction
of some form of diffusivity and viscosity, the proper relaxation of a simulation to compute reliable statistics,
the role of boundary conditions, and a list of criteria to judge on the amount of uncertainty introduced
by the different modelling approaches to convection. Affordability issues related to numerical simulations
are summarized before a specific topic, convective overshooting, is selected as a process which in 
Sect.~\ref{sec-5} serves as an example for the application of the different modelling principles introduced
in the preceding sections. The fundamental non-locality of overshooting is demonstrated using a semi-analytical 
approach, the Reynolds stress formalism. Its extension by moment budgets computed from numerical simulations
is discussed before the DA white dwarfs are introduced as a highly convenient class of stellar
objects for the study of overshooting. The necessary preconditions for a solid study of this process
are then explained including the constraints put by length and time scales and by several dimensionless
quantities, in particular the P\'eclet number. Finally, the main results on studying overshooting in DA white dwarfs 
as published when writing this introduction are presented. The concepts presented in this text, while developed here 
also for rather specific applications, are nevertheless of general interest in studies of convection in both 
astrophysics and geophysics. They are hoped to be useful also for approaching more advanced subjects related 
to convection in further detail.

\vspace{1cm}

\begin{acknowledgement}
Support of the author by the Austrian Science Fund (FWF), project P29172-N27, is gratefully acknowledged.
\end{acknowledgement}

%

\bibliography{lrr-stellar_convection-refs,lrr-stellar_convection-refs_part2}

%
%
%
%

\end{document}